\documentclass[apj]{emulateapj}

\usepackage{txfonts}
\usepackage{natbib}
\newcommand{\asec} {\mbox{$^{\prime \prime}$} }
\newcommand{\amin} {\mbox{$^{\prime}$}}
\usepackage{mathrsfs}
\usepackage{amsfonts}
\usepackage{enumerate}
\bibpunct{(}{)}{;}{a}{}{,}

\newdimen\digitwidth    
\setbox1=\hbox{0}       
\digitwidth=\wd1        
\catcode`"=\active      

\def\gsimeq{\hbox{\raise0.5ex\hbox{$>\lower1.06ex\hbox{$\kern-1.07em{\sim}$}$}}} 
\def\lsimeq{\hbox{\raise0.5ex\hbox{$<\lower1.06ex\hbox{$\kern-1.07em{\sim}$}$}}} 

\slugcomment{July 15, 2014}
\shorttitle{ \textsc{NGC4945}}
\shortauthors{Puccetti et al.}

\begin{document}

\title{The variable hard X--ray emission of  \textsc{NGC4945} as observed by {\it NuSTAR}}

\author{Simonetta Puccetti\altaffilmark{1,2}, Andrea Comastri
  \altaffilmark{3}, Fabrizio Fiore \altaffilmark{2}, Patricia
  Ar{\'e}valo \altaffilmark{4,5}, Guido Risaliti \altaffilmark{6,7},
  Franz E. Bauer \altaffilmark{4,8}, William N. Brandt
  \altaffilmark{9,10}, Daniel Stern \altaffilmark{11}, Fiona
  A. Harrison \altaffilmark{12}, David M. Alexander\altaffilmark{13},
  Steve E. Boggs \altaffilmark{14}, Finn E. Christensen
  \altaffilmark{15}, William W. Craig \altaffilmark{16,14}, Poshak
  Gandhi \altaffilmark{13}, Charles J. Hailey \altaffilmark{17},
  Michael R. Koss \altaffilmark{18}, George B. Lansbury
  \altaffilmark{13}, Bin Luo\altaffilmark{9,10}, Greg M. Madejski
  \altaffilmark{19}, Giorgio Matt \altaffilmark{20}, Dominic
  J. Walton\altaffilmark{12}, Will Zhang \altaffilmark{21} }

\altaffiltext{1}{ASDC--ASI, Via del Politecnico, 00133 Roma, Italy}
\altaffiltext{2}{INAF--Osservatorio Astronomico di Roma, via Frascati
  33, 00040 Monte Porzio Catone (RM), Italy}
\altaffiltext{3}{INAF--Osservatorio Astronomico di Bologna, via Ranzani
1, 40127 Bologna, Italy}
\altaffiltext{4}{Instituto de Astrof\'{\i}sica, Facultad de F\'{i}sica, Pontificia Universidad Cat\'{o}lica de Chile, 306, Santiago 22, Chile}
\altaffiltext{5}{Instituto de F\'{i}sica y Astronom\'{i}a, Facultad de Ciencias, Universidad de Valpara \'{i}so, Gran Bretana N 1111, Playa Ancha, Valpara\'iso, Chile}
\altaffiltext{6}{INAF--Osservatorio Astrofisico di Arcetri, Largo
  E. Fermi 5, 50125 Firenze, Italy}
\altaffiltext{7}{Harvard--Smithsonian Center for Astrophysics, 60 Garden Street, Cambridge, MA 02138, USA} 
\altaffiltext{8}{Space Science Institute, 4750 Walnut Street, Suite 205, Boulder, Colorado 80301}
\altaffiltext{9}{Department of Astronomy and Astrophysics, The Pennsylvania State University, 525 Davey Lab, University Park, PA 16802} 
\altaffiltext{10}{Institute for Gravitation and the Cosmos, The Pennsylvania State University, University Park, PA 16802}
\altaffiltext{11}{Jet Propulsion Laboratory, California Institute of Technology, 4800
Oak Grove Drive, Pasadena, CA 91109, USA}
\altaffiltext{12}{Cahill Center for Astronomy and Astrophysics, California Institute of
Technology, Pasadena, CA, 91125 USA}
\altaffiltext{13}{Department of Physics, Durham University, Durham DH1 3LE, UK}
\altaffiltext{14}{Space Sciences Laboratory, University of California, Berkeley, CA 94720}
\altaffiltext{15}{DTU Space -- National Space Institute, Technical University of Denmark, Elektrovej 327, 2800 Lyngby, Denmark}
\altaffiltext{16}{Lawrence Livermore National Laboratory, Livermore, CA 945503}
\altaffiltext{17}{Columbia Astrophysics Laboratory, Columbia University, New York, NY 10027}
\altaffiltext{18}{Institute for Astronomy, Department of Physics, ETH Zurich, Wolfgang--Pauli--Strasse 27, CH--8093 Zurich, Switzerland}
\altaffiltext{19}{Kavli Institute for Particle Astrophysics and Cosmology, SLAC National Accelerator Laboratory, Menlo Park, CA 94025}
\altaffiltext{20}{Dipartimento di Matematica e Fisica, Universit`a Roma Tre, via della Vasca Navale 84, 00146 Roma, Italy}
\altaffiltext{21}{NASA Goddard Space Flight Center, Greenbelt, MD 20771}

\begin{abstract}

We present a broadband ($\sim$0.5 -- 79~keV) spectral and temporal
analysis of multiple {\it NuSTAR} observations combined with archival
{\it Suzaku} and {\it Chandra} data of \textsc{NGC4945}, the brightest
extragalactic source at 100 keV. We observe hard X--ray ($> 10$ keV)
flux and spectral variability, with flux variations of a factor 2 on
timescales of 20~ksec.  A variable primary continuum dominates the
high energy spectrum ($>10$ keV) in all the states, while the
reflected/scattered flux which dominates at E$< 10$~keV stays
approximately constant. From modelling the complex
reflection/transmission spectrum we derive a Compton depth along the
line of sight of $\tau_{\rm Thomson}\sim 2.9$, and a global covering
factor for the circumnuclear gas of $\sim 0.15$. This agrees with the
constraints derived from the high energy variability, which implies
that most of the high energy flux is transmitted, rather that
Compton--scattered. This demonstrates the effectiveness of spectral
analysis in constraining the geometric properties of the circumnuclear
gas, and validates similar methods used for analyzing the spectra of
other bright, Compton--thick AGN. The lower limits on the e--folding
energy are between $200-300$ keV, consistent with previous {\it
  BeppoSAX}, {\it Suzaku} and {\it Swift} BAT observations. The
accretion rate, estimated from the X--ray luminosity and assuming a
bolometric correction typical of type 2 AGN, is in the range $\sim
0.1-0.3$ $\lambda_{\rm Edd}$ depending on the flux state.  The
substantial observed X--ray luminosity variability of \textsc{NGC4945}
implies that large errors can arise from using single--epoch X--ray
data to derive $L/L_{\rm Edd}$ values for obscured AGNs.

\end{abstract}


\maketitle

\section{Introduction}

The almost edge-on spiral starburst galaxy \textsc{NGC4945} hosts one
of the nearest AGN ($D\sim 3.8$ Mpc, see e.g. Mouhcine et al. 2005,
Mould \& Sakai 2008, Tully et al. 2008, Jacobs et al. 2009, Tully et
al. 2009, Nasonova et al. 2011). It has been the subject of many
observations spanning a broad X--ray band, from sub--keV to hundreds of
keV energies. X--ray emission from the infrared galaxy \textsc{NGC4945} was
discovered by {\it Ginga} and identified as a deeply buried AGN
(Iwasawa et al. 1993). It is the brightest Seyfert 2 galaxy in the
very hard ($\sim 50-100$ keV) X--ray sky, with a $10-50$ keV flux of
the order of 10$^{-10}$ erg cm$^{-2}$ s$^{-1}$, as seen by the {\it
  Compton Gamma-ray Observatory} (Done, Madejski, and Smith 1996), the
{\it Rossi Timing X--ray Explorer} (Madejski et al. 2000) and the PDS
instrument onboard {\it BeppoSAX} (Guainazzi et al. 2000).

The broad--band spectrum, using high--energy data extending up to
100~keV in conjunction with softer X-ray observations from {\it ASCA}
and {\it Ginga}, is described by a power--law continuum with strong
photoelectric absorption (i.e. Done, Madejski, Smith 1996), with a
column density $N_{\rm H}$ (under the assumption of Solar abundances)
of about $4 \times 10^{24}$ cm$^{-2}$. This column and the observed
high energy variability identify \textsc{NGC4945} as a
transmission-dominated Compton--thick AGN. In such sources, the
high-energy (E$>10$~keV) primary continuum penetrates the obscuring
matter and peaks at $15-20$ keV, depending on the actual value of the
optical depth.  Compton--thick AGN are predicted to provide a
significant contribution (from 10 to 30\%, depending on the specific
model assumptions) to the cosmic X--ray background spectrum at
energies close to its $\sim 20-30$ keV peak (Gilli et al. 2007;
Treister et al. 2009). The X--ray spectrum of \textsc{NGC4945} also
shows a strong iron K$_\alpha$ line (EW $> $1 keV) that is thought to
arise from fluorescence in matter which is more remote from the
nucleus than the primary continuum (e.g. Matt et al. 1996).

The hard X--ray emission in the {\it BeppoSAX} $15-200$ keV band has
been observed to vary by a factor of two on timescales as short as 10$^4$~s
(Guainazzi et al. 2000). This suggests that the solid angle subtended
by the Compton--thick absorber, as seen by the source, is unlikely to
be large, and the heavy absorption of the primary continuum is
probably confined to a disk-like structure or a ``skinny'' torus with
a relatively small half--opening angle ($<10^{\circ}$), otherwise the
scattered X--rays would noticeably dilute the hard X--ray variability
(Madejski et al. 2000). \textsc{NGC4945} was also observed with {\it
  Suzaku} (Itoh et al. 2008), which confirmed the strong hard X-ray
variability of the heavily obscured primary continuum. The spectrum
below 10 keV is dominated by a constant reflected continuum along with
neutral iron K$_\alpha$, K$_\beta$ and Nickel K$_\alpha$ lines. An
extensive discussion of the origin of the hard X--ray emission from
\textsc{NGC4945} is presented in Yaqoob (2012) using
state--of--the--art Monte Carlo simulations (the \textsc{mytorus}
model) to treat absorption and scattering in the Compton--thick
regime. They conclude, in agreement with previous studies, though on much
more solid statistical and physical grounds, that a clumpy medium with
a small covering factor provides the best description of the observed
spectrum and variability.

\textsc{NGC4945} has been observed by many instruments in the $\sim
0.5-10$ keV band, and all data obtained so far indicate that it is
almost constant in this energy range. High angular resolution
observations with {\it Chandra} (Schurch et al. 2002, Done et
al. 2003) reveal a clear change in the nature of the $\sim 0.5-7$ keV
X--ray spectrum with distance from the nucleus. The compact,
unresolved emission confirms the basic absorbed power--law scenario
given above. The closest ($< 50-100$ pc) soft X--ray ($\sim 0.5-5$
keV) emission is likely associated with strong circumnuclear starburst
activity that is well modelled by a two temperature ($\sim 0.9$ and
$\sim 6$ keV) optically thin thermal plasma (Schurch et al. 2002). The
starburst region also contains cold reflection signatures (see also
Marinucci et al. 2012). Despite the fact that the X--ray images in the
line emission band ($6.2-6.7$ keV) and in the $2-7$ keV band show an
asymmetric and clumpy spatial distribution with a size of tens of
parsecs, they almost perfectly match and are likely to be associated
with the cold reflector. Unresolved clumpiness may explain the low
covering factor inferred from the hard X--ray variability. The more
distant region (beyond $\sim 50-100$ pc) reveals distinctly different
emission, with the spectrum showing soft continuum plus emission
lines. This more extended emission appears morphologically as a plume,
and was modelled as a combination of photoionized and collisionally
ionized plasma, with $kT \sim 0.7$ keV (see Done et al. 2003 for
details). The galaxy also contains off--nuclear variable ULXs
(e.g. Brandt et al. 1996, Swartz et al. 2004, Isobe et al. 2008,
Walton et al. 2011).

The nucleus of \textsc{NGC4945} is a strong source of megamaser
activity, detected in H$_{\rm 2}$O (Dos Santos \& Lepine 1979).
Detailed, velocity--resolved maps reveal multiple megamaser spots
revolving around the central black hole: this allows the precise
determination of the black hole mass, at $1.4 \times 10^{6} M_{\odot}$
(Greenhill, Moran, \& Herrnstein 1997). The megamaser spots are
irregularly distributed and clumpy.

Madejski et al. (2000) and Done et al. (2003) suggested that the
megamaser spots are aligned with the absorbing disk/torus at $0.1-1$
pc. In addition to this dense absorbing disk, the nucleus is
completely embedded in a column of $\sim 10^{23}$ cm$^{-2}$ at
$\sim 25$ pc, which may be due to a bar--driven gas inflow (Ott et
al. 2001) or perhaps a high--latitude extension of the disk/torus. The
extreme absorption region is surrounded by an extended, dusty, lower
absorption region on scales of $50-100$ pc in the starburst region.

The nucleus of \textsc{NGC4945} is characterized by a high and
variable accretion rate. The Eddington ratio ($\lambda_{\rm Edd}
\equiv L_{\rm BOL}/L_{\rm EDD}$) is reported to be in the range
0.1--0.5 (Guainazzi et al. 2000, Madejski et al. 2000; Done et
al. 2003; Itoh et al. 2008). Yaqoob (2012) reported values as high as
$\lambda_{\rm Edd} \simeq 2$.  It is however important to note that
super Eddington accretion rates are obtained assuming a distance of
about 8 Mpc, which is about twice the distance commonly adopted for
\textsc{NGC4945} ($D\sim 3.8$ Mpc).

\textsc{NGC4945} is the only known Compton thick Seyfert 2 galaxy in
the nearby Universe (i.e. $z<0.004$) that shows rapid variability in
the very hard $>$ 10 keV X--rays. A detailed state--resolved spectral
analysis with a very short time binning, which could not performed
with previous hard X--ray data, is the main scientific focus of the
medium--deep monitoring {\it NuSTAR} observations described in this
paper.

In \S~2, we describe the {\it NuSTAR}
observations and archival observations. In \S~3 and 4 we detail
the variability properties. \S~5 presents the state--resolved
spectral analysis. The results are discussed in \S~6 and the
conclusions comprise \S~7.

Through out this paper, we adopt a \textsc{NGC4945} distance of $3.8$ Mpc.

\section{X--ray observations and data reduction}

\subsection{{\it NuSTAR}}

We analyzed three {\it NuSTAR} observations of \textsc{NGC4945}
performed in February, June and July 2013 with the two focal plane modules A
(FPMA) and B (FPMB) (Harrison et al. 2013), for a total exposure time
of $\sim 150$ ksec. Table~\ref{tablog} gives a log of the 
{\it NuSTAR} observations.

\begin{table*}
\footnotesize
\caption{{\it NuSTAR}  \textsc{NGC4945} observation log}
\centering
\begin{tabular}{lcccccccc}
\hline
Observation ID$^a$ & RA\_PNT$^b$ & DEC\_PNT$^c$ &  Exposure$^d$ & Start Date$^e$ & rate$^f$& background$^g$ \\
 & (deg.) & (deg.) & (ksec) &  &  (cts/s) & \\
\hline
 60002051002 & 196.406  & -49.4605 & 45.1  & 2013-02-10T22:31:07 & 0.462$\pm$0.002 & 3.9\%\\
 60002051004 & 196.3271 & -49.461  & 54.5 & 2013-06-15T04:56:07 & 0.519$\pm$0.002 & 3.3\%\\
 60002051006 & 196.3334 & -49.4907 & 50.5  & 2013-07-05T22:51:07 & 0.349$\pm$0.002 & 5.9\%\\
\hline 
\end{tabular}

$^a$Observation identification number; $^b$right Ascension of the
pointing; $^c$declination of the pointing; $^d$ total net exposure
time; $^e$start date and time of the observation;
$^f$mean value of the net count rate in the circular source extraction
region with 75$\asec$ radius in the energy range $3-79$ keV; $^g$
background percentage in the circular source extraction region with
75$\asec$ radius and in the energy range $3-79$ keV.
\label{tablog}
\end{table*}

The telemetry raw fits files were processed using the {\it NuSTAR}
Data Analysis Software package v. 1.3.1 (NuSTARDAS: Perri et
al. 2013). Calibrated and cleaned event files were produced using the
calibration files in the {\it NuSTAR} CALDB (20131223) and standard
filtering criteria with the {\it nupipeline} task. For the July
observation we followed a different procedure because portions of the
observation had the raw positions of the laser spots outside of the
calibrated range of the position sensing detector (see Harrison et
al. 2013). Therefore those data would be not linearized using the
standard calibration metrology grid file (see Perri et al. 2013). As a
consequence, the sky position of the X--ray photons would be not calculated
by the standard pipeline setting and these time intervals would be excluded
from the calibrated and cleaned data. However, for the analysis of
bright sources, the lost exposure time can be recovered by using the
raw coordinates of the laser spots without applying the calibration
correction; this is done at the expense of some positional
uncertainty.

\begin{figure}
\begin{center}
\includegraphics[width=5.8cm]{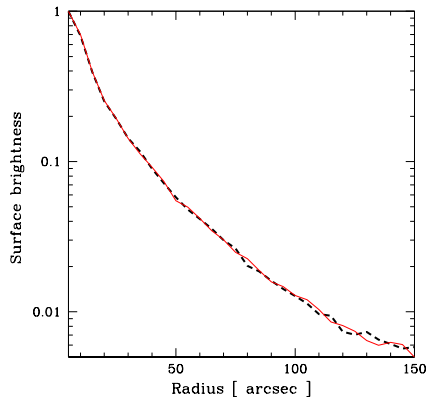}
\caption{$3-79$ keV source brightness profiles for the July
  observation. The red solid profile is the result obtained using the cleaned
  event file produced with the standard analysis; the black dashed
  profile is the result of the non--standard analysis.}
\label{bri}
\end{center}
\end{figure}

We applied this non--standard analysis to the July observation; the
reliability of the reconstruction of the sky coordinates was tested by
comparing the two source brightness profiles generated using the
cleaned event file produced with the standard and non--standard
analysis. We verified that the two profiles are fully consistent
($99.9\%$, see Fig. \ref{bri}). Moreover, looking at the difference
between the two brightness profiles, most of the photons are caught in
the source extraction region (radius of 75$\asec$, see
below). Therefore, we safely used the calibrated and cleaned data
generated following the non--standard analysis.

\begin{figure}
\begin{center}
\includegraphics[width=6.5cm]{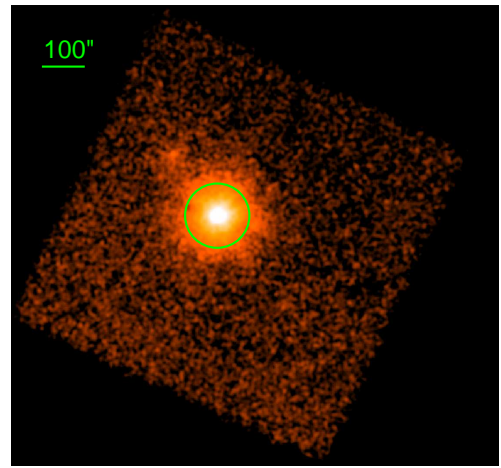}
\caption{ $3-79$ keV {\it NuSTAR} image for the June observation
    and FPMA module. The image was smoothed with a Gaussian filter
    with $\sigma=1.5$. The green circle is centered on the peak of the
    emission and has a radius of 75$\asec$.}
\label{imanu}
\end{center}
\end{figure}

We extracted the {\it NuSTAR} source and background spectra using the
{\it nuproducts} task included in the NuSTARDAS package using the
appropriate response and ancillary files. We extracted spectra and
light curves in each focal plane module (FPMA and FPMB) using circular
apertures of radius 75$\asec$ centered on the peak of the emission in
the $3-79$~keV data (see Fig. \ref{imanu}). This choice maximizes the
S/N of the data. Background spectra were extracted using source--free
regions on the same detectors where the source is detected. As shown
in Tables~1 and 4, the background count--rates are a low fraction ($<
10\%$) of the source count--rates for most observations in most energy
intervals.

The large number of counts in the {\it NuSTAR} \textsc{NGC4945}
spectra allows statistical grouping to oversample the instrument
resolution. However, this can cause problems during spectral fitting
because the bins are not completely independent. For this reason, the
spectra were binned according to two criteria: (i) following the
energy resolution multiplied by a factor $\sim 0.4$ at all energies,
when possible; (ii) to have a signal--to--noise ratio $> 4.5$.

For the spectral analysis, in all cases we co--added the spectra of
each focal plane module into a single spectrum (e.g. for spectra at
different time intervals), and we also combined the corresponding
background spectra, response and ancillary files. We used the {\it
  addascaspec} FTOOLS v.6.13, which combines spectra according to the
method explained in the {\it ASCA} ABC
guide\footnote{http://heasarc.gsfc.nasa.gov/docs/asca/abc/}. The
background normalization is calculated as for {\it ASCA}
data\footnote{
  http://heasarc.gsfc.nasa.gov/docs/asca/abc$\_$backscal.html}.

\subsection{ {\it Chandra}}

The angular resolution of {\it NuSTAR} is dominated by the optics, and
has an 18$\asec$ FWHM with a half--power diameter of 58$\asec$ (Harrison
et al. 2013). Therefore, {\it Chandra}, with its unsurpassed X--ray
angular resolution (FWHM$\sim 0.5 \asec$), is an excellent complement
to {\it NuSTAR} for analyze the nucleus and nearby regions, and
for studying the possible contamination from unresolved sources.

We analyzed three archival {\it Chandra} observations collected in
2000-2004 (see Table~\ref{tabarchive}) using the {\it Chandra}
Interactive Analysis Observations (CIAO) software (v4.5; Fruscione et
al. 2006) and the standard data reduction procedures. After cleaning
for background flaring the total exposures are $\sim 33$ ksec,
  $\sim 76$ ksec and $\sim 97$ ksec for the 2000 and the two 2004
  HETG--zeroth observations, respectively. The {\it specextract} task
was used to extract the spectra; the 2004 spectra were combined using
{\it addascaspec} from FTOOLS v.6.13, as described in \S~2.1. The
spectra were binned to have at least 30 total counts per bin.

\begin{figure}
\begin{center}
\includegraphics[width=7cm]{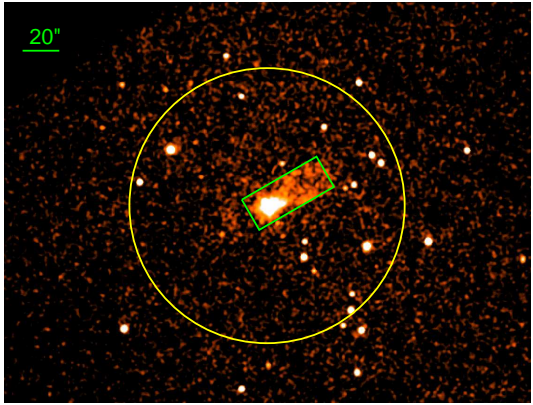}
\caption{$0.5-7$ keV image mosaic of the three {\it Chandra}
  observations (see Table~\ref{tabarchive}). The green 47$\asec$
  $\times$ 19$\asec$ box marks the region used to extract the {\it
    Chandra} spectra, which includes the nuclear source plus a
  spectrally soft, conically shaped X--ray plume (Schurch et al. 2002,
  Done et al. 2003). The yellow circle (75$\asec$ radius) marks the
    {\it NuSTAR} source extraction region.}
\label{chandraima}
\end{center}
\end{figure}

The {\it Chandra} spectra were extracted in a 47$\asec$ $\times$
19$\asec$ box centered on R.A.$=$13:05:26.331, Dec.$=-$49:27:56.41 (see
Fig. \ref{chandraima}) to include the nuclear source plus a spectrally
soft, conically shaped X--ray plume, due to starburst/superwind
  gas in multiphase state, which extends 30$\asec$ ($\sim 500$ pc) to
the northwest (Schurch et al. 2002, Done et al. 2003) and is not
resolved by {\it NuSTAR}.

\begin{table*}
\footnotesize
\caption{ \textsc{NGC4945}  archival data}
\centering
\begin{tabular}{lccccccccc}
\hline
Observatory & Observation ID$^a$ &  Exposure$^b$ & Start Date$^c$ & ref$^d$\\
 & & (ksec) & \\
\hline
{\it BeppoSAX} & 50809001 & 40 & 1999-07-01 04:43:13.2 & Guainazzi et al. 2000 \\
 {\it Chandra} ACIS-S  & 864 & 50 & 2000-01-27 19:00:10 & Schurch et al. 2002 \\
 {\it Chandra} HETG ACIS--S & 4899 & 79 & 2004-05-28 04:47:04 & Marinucci et al. 2012 \\
 {\it Chandra} HETG ACIS--S & 4900 & 97 & 2004-05-29 18:36:22 & Marinucci et al. 2012\\
 {\it Suzaku} & 705047010 & 39 & 2010-07-04 15:15:58 & Marinucci et al. 2012 \\
 {\it Suzaku} & 705047020	& 44 & 2010-07-09 23:50:49 &  Marinucci et al. 2012 \\
 {\it Suzaku} & 705047030	 & 40 & 2010-07-26 01:36:35 &  Marinucci et al. 2012 \\
 {\it Suzaku} & 705047040 & 39 & 2010-08-30 21:34:36 &  Marinucci et al. 2012 \\
 {\it Suzaku} & 705047050	& 46 & 2011-01-29 02:05:50 &  Marinucci et al. 20122 \\
{\it Swift} BAT & 70 month catalog  & 7811$^e$   & & Baumgartner et al. 2013 \\
\hline 
\end{tabular}
  
$^a$ Observation identification number; $^b$ total net exposure time;
$^c$ start date and time of the observation interval; $^d$ reference;
$^e$ on--axis equivalent exposure.
\label{tabarchive}
\end{table*} 

About 13 serendipitous point sources with S/N $\ge 3$ in the $2-7$ keV
band are identified within the area corresponding to the {\it NuSTAR}
extraction radius (75$\asec$), and about 20 sources are within the
{\it Suzaku} spectral extraction region (see next section). By
comparing the X--ray spectra of the nuclear region (including the
plume), with the spectra of the serendipitous point sources we
estimate that the contamination by the nearby bright sources is of the
order of $\sim 60\%$ in the $4-6$ keV energy band in both {\it
    NuSTAR} and {\it Suzaku} spectra (see Table~\ref{tabfac}). This
  evaluation is approximate and for guidance only. The serendipitous
  sources could be variable and their exact contribution depends on
  the energy range and detector PSF.

There is no evidence of any iron line emission in their spectra. The
co--added spectrum of the five brightest ``contaminants'' is
best--fitted by an absorbed power--law ($\Gamma=$2.1$\pm_{0.1}^{0.2}$,
$N_{\rm H}$$=$1.1$\pm_{0.3}^{0.4} \times 10^{22}$ cm$^{-2}$) plus a
low temperature thermal component ($kT$$=$0.22$\pm{0.07}$ keV).
 
\begin{table}
\footnotesize
\centering
\caption{Chandra serendipitous source contribution}
\begin{tabular}{cccc}
\hline
Energy Range & Fac$^a$ & Fac$^b$ \\
 keV & & & \\
\hline
0.5-2 & 82\% & 86\% \\
2-3   & 85\% & 88\% \\
3-4   & 74\% & 77\% \\
4-5   & 69\% & 72\% \\
5-6   & 47\% & 55\% \\
6-7   & 18\% & 18\% \\
\hline 
\end{tabular}

$^a$ The contribution of Chandra serendipitous sources to the
  \textsc{NGC4945} spectrum. The {\it NuSTAR} spectral
  extraction source region is a circle with 75$\asec$ radius;
  $^b$ same as $^a$ for the {\it Suzaku} spectral extraction source region
  (i.e. circle of 1.85$\amin$ radius).
\label{tabfac}
\end{table}

\subsection{ {\it Suzaku}}

The energy resolution of {\it Suzaku} at $\sim 6$ keV and the good
counting statistics of the archival observations allow us to much
better constrain the spectral shape and the intensity of the iron line
complex.  Therefore, in the spectral analysis we considered five {\it
  Suzaku} observations performed between July 2010 and January 2011
(see Table~\ref{tabarchive}). The X--ray imaging Spectrometer (XIS)
data were extracted from a circular region with radius of 1.85$\amin$
(see Marinucci et al. 2012, for details on data reduction). The
  spectral cross--calibration between {\it NuSTAR} and {\it Suzaku}
  XIS is fairly good, with fluxes consistent within $\sim10\%$ (Walton
  et al. 2014, Brenneman et al. 2014).

The HXD--PIN data were reduced following the {\it Suzaku} data
reduction guide (the ABC guide Version 5.0) using the rev2 data,
which include all four cluster units. We used the instrumental
background event file provided by the HXD--PIN team (NXB; Kokubun et
al. 2007), which has systematic uncertainty of $\pm(3-5)\%$ (at the
1$\sigma$ level). We extracted the source and background light curve
in the same time interval, and corrected the source light curve for
the detector dead time. The cleaned net exposure times were $\sim 32$,
37, 32, 33 and 50 ksec for the five {\it Suzaku} observations,
respectively. The cosmic X--ray background spectrum was simulated
using the spectral shape in Boldt (1987) and Gruber et al. (1999), and
then added to the instrumental background, assuming a constant light
curve.

The HXD PIN count rate varies by a factor of $\sim 3.7$ in the $16-80$
keV energy range over the five observations.

\subsection{{\it BeppoSAX} PDS}

\textsc{NGC4945} was observed by {\it BeppoSAX} in 1999 (see
Table~\ref{tabarchive}). The PDS (Phoswitch Detector System, Frontera
et al. 1997) data were calibrated and cleaned using the SAXDAS
software with the standard method ``fixed rise time threshold'' for
background rejection. PDS light curves are well known to exhibit
spikes on timescales between a fraction of a second to a few seconds
and usually most counts from spikes are recorded below 30 keV. To
screen the PDS data for such spikes we followed the method suggested
in the NFI user guide (Fiore et al. 1999). The PDS count rate light
curve varies by a factor $\sim 1.5$ over a time scale of a few 10$^4$
sec (see Guainazzi et al. 2000).

\subsection{{\it Swift} BAT}

We retrieve the {\it Swift} BAT (Gehrels et al. 2004) \textsc{NGC4945}
light curve and spectrum from the last public stack archive (see
Table~\ref{tabarchive}). The on--axis integration time is $\sim 7.81$
Msec and the source has S/N $\sim 80$ in the $14-195$ keV energy
range. The count rate uncertainty is fairly constant up to 100 keV; at
higher energies the noise dominates. The BAT $14-195$ count rate light
curve varies by a factor $\sim 3$ over the 70 months.

\section{{\it NuSTAR} light curves}

\begin{figure*}
\begin{tabular}{cc}
\centering
\includegraphics[width=9.4cm]{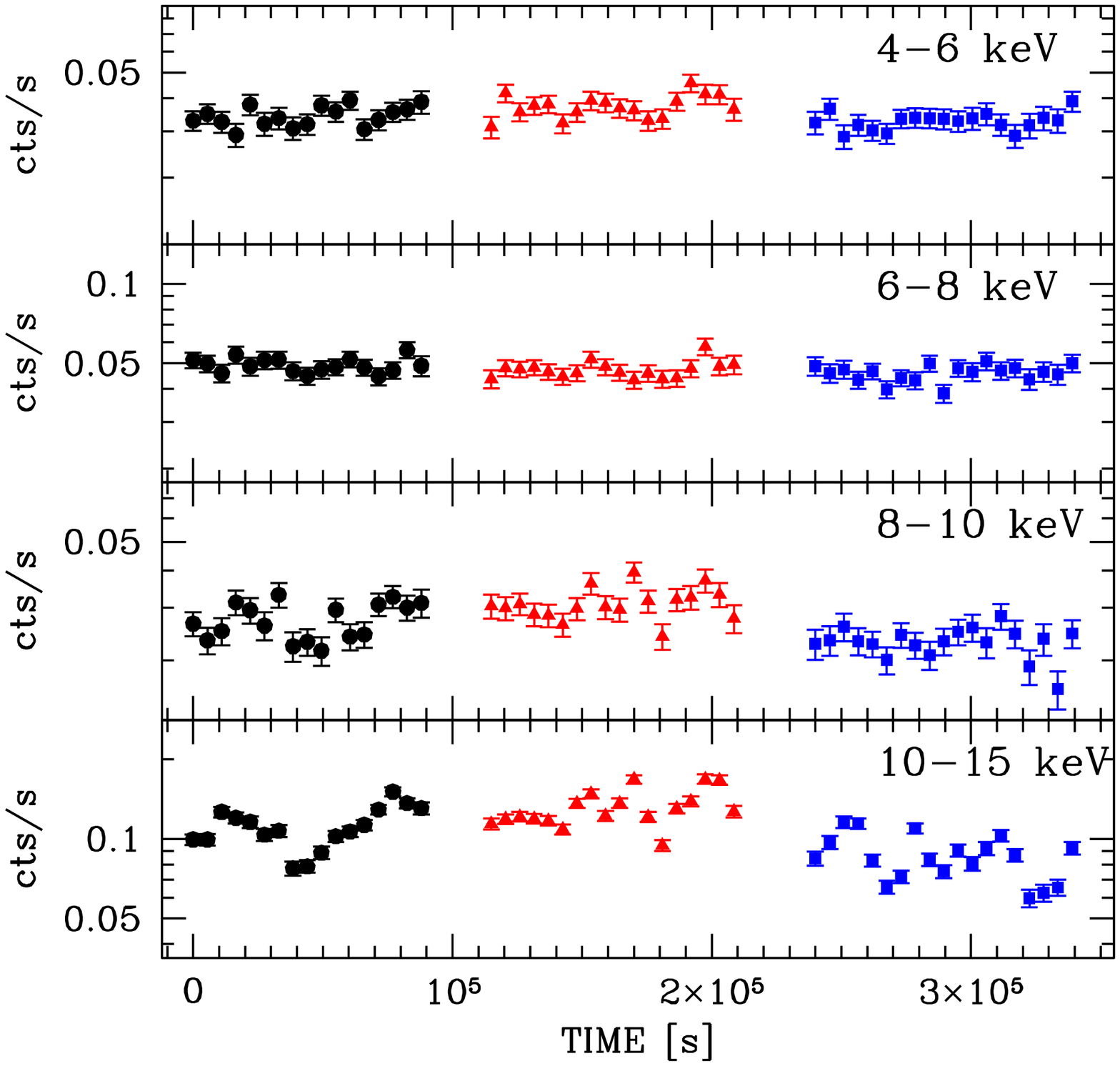}
\includegraphics[width=9.4cm]{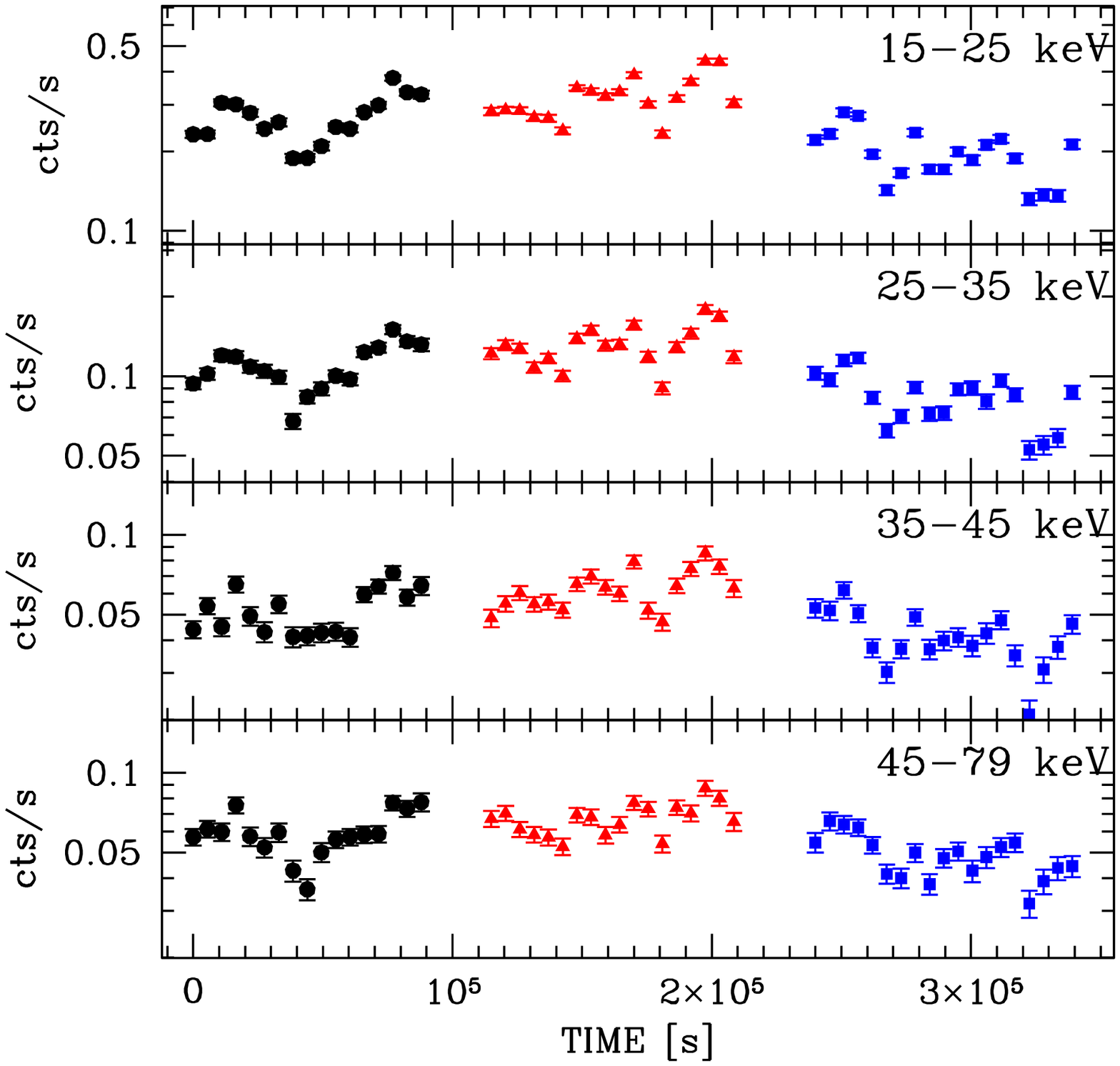}
\end{tabular}
\caption{Light curves in bins of 5500 s ($\sim 1$ satellite
  orbit). The count rates are the mean values between those detected
  by the FPMA and FPMB modules, corrected for livetime, PSF losses and
  vignetting, but not for background. {\it Left panel}, from top to
  bottom: $4-6$ keV, $6-8$ keV, $8-10$ keV and $10-15$ keV count
  rates. {\it Right panel}, from top to bottom: $15-25$ keV, $25-35$
  keV, $35-45$ keV and $45-79$ keV count rates. Black dots, red
  triangles and blue squares refer to observations in February, June
  and July, respectively. The real time interval between black dots
  and red triangles is about 4 months, whereas between red triangles
  and blue squares is about 1 month. The y$-$axis dynamic range
    is 0.9 in logarithmic scale, in each panel.}
\label{lcurves}
\end{figure*}

The three {\it NuSTAR} observations span a timescale of $\sim 50$ ksec
each (see Table~\ref{tablog}). Fig. \ref{lcurves} shows the {\it
  NuSTAR} light curves in bins of 5500 s ($\sim 1$ satellite orbit) in
eight energy bands ($4-6$, $6-8$, $8-10$, $10-15$, $15-25$, $25-35$,
$35-45$ and $45-79$ keV); the light curves were corrected for
livetime, PSF losses, vignetting, but not for background, which is
negligible (see Table~\ref{tabfaclc}). Note that the y$-$axis dynamic
range is the same in each panel of Fig. \ref{lcurves}.

We used the $\chi^2$ test with a minimum confidence level of 2\% to
assess the variability of the light curves. We found that the $4-6$
keV and $6-8$ keV light curves are consistent with being constant and
are not correlated with the other six energy bands.  The $4-6$ and
$6-8$ keV count rates are also constant among the three observations
(taken within a time interval of $\sim 5$ months). The $8-10$~keV
light curves are consistent with being constant only in the July
observation. The $8-10$~keV variability pattern of the February and
June observations correlates with all the observing periods in the
harder ($> 10$ keV) bands.  The $10-15$, $15-25$, $25-35$, $35-45$ and
$45-79$ keV light curves are not constant at 99.9\% confidence level,
and are strongly correlated with each other in all the observations
(see Fig.~\ref{lcurves} and the Spearman rank correlation coefficients
in Table~\ref{tabfaclc}).

\begin{table*}
\footnotesize
\caption{Light curve mean count rate and correlation coefficient}
\centering
\begin{tabular}{llllllllll}
\hline
Energy Range & Feb$^a$ & back\_Feb$^b$& Jun$^c$ &  back\_Jun$^d$ & Jul$^e$ &  back\_Jul$^f$ & r,P (Feb)$^g$ & r,P (Jun)$^h$ & r,P (Jul)$^i$ \\
(keV) & (cts/s) & & (cts/s) &  & (cts/s) & & & & \\
\hline
4-6  & 0.034$\pm$0.003  & $\sim$8\%   &   0.037$\pm$0.004  & $\sim$6.5\% &  0.033$\pm$0.002  &  $\sim$9\% &  0.17, 50\%    & 0.53, $3\%$  & 0.04, $>$50\%  \\
6-8  & 0.049$\pm$0.003 &  $\sim$4\%    &  0.047$\pm$0.004   & $\sim$4\% &   0.046$\pm$0.003  &  $\sim$5\% &  0.14, $>$50\% & 0.38, $>10\%$ & 0.26, 20\%    \\
8-10  & 0.027$\pm$0.004 &  $\sim$6\%   &  0.031$\pm$0.004   & $\sim$4\% &   0.023$\pm$0.003  &  $\sim$7\% &  0.75, $<$0.1\%  & 0.77, $<$0.1\% & 0.44, 5\%     \\
10-15 & 0.11$\pm$0.02  &  $\sim$2.5\% &  0.13$\pm$0.02     & $\sim$2\% &   0.09$\pm$0.02    &  $\sim$3\% &  0.98, $<$0.1\%  & 0.96, $<$0.1\% & 0.96, $<$0.1\%  \\
15-25 & 0.27$\pm$ 0.05 & $\sim$1.5\% & 0.32$\pm$0.06 & $\sim$1\% &  0.19$\pm$0.04 &  $\sim$2.5\% & 1, 0\%           &  1, 0\%            &   1, 0\%               \\   
25-35 & 0.11$\pm$ 0.02 & $\sim$4.5\% & 0.13$\pm$0.02 & $\sim$4\% &  0.08$\pm$0.02 &  $\sim$6.5\% & 0.95, $<$0.1\% &  0.94, $<$0.1\% &    0.94, $<$0.1\%    \\  
35-45 & 0.05$\pm$ 0.01 & $\sim$5\%   & 0.06$\pm$0.01 & $\sim$4\% &  0.04$\pm$0.01 &  $\sim$8\%   & 0.82, $<$0.1\% &  0.91, $<$0.1\% &    0.89, $<$0.1\%    \\  
45-79 & 0.06$\pm$ 0.01 & $\sim$13\%  & 0.07$\pm$0.01 & $\sim$12\% & 0.05$\pm$0.01 &  $\sim$19\%  & 0.84, $<$0.1\% &  0.77, $<$0.1\% &    0.81, $<$0.1\%  \\    

\hline
\end{tabular}

$^a$ Mean values of the light curve count rates for the 2013 February
observation, in the energy range indicated in the first column of the
table. $^b$ Background percentage in the circular source extraction
region with 75$\asec$ radius; $^c$ same as $^a$ for the 2013 June
observation; $^d$ same as $^b$ for the 2013 June observation; $^e$
same as $^a$ for the 2013 July observation; $^f$ same as $^b$ for the
2013 July observation; $^g$ r is the Spearman rank correlation
coefficient between the light curves of the 2013 February observation,
in the energy range indicated in the first column of the table and the
energy range $15-25$ keV; P is the probability that the correlation is
not statistically significant, evaluated by the Student's t test. $^h$
same as $^g$ for the 2013 June observation; $^i$ same as $^g$ for the
2013 July observation. The quoted errors are at the 90\% confidence
level for one parameter of interest.

\label{tabfaclc}
\end{table*} 

\begin{figure*}
\begin{tabular}{cc}
\centering
\includegraphics[width=9.4cm]{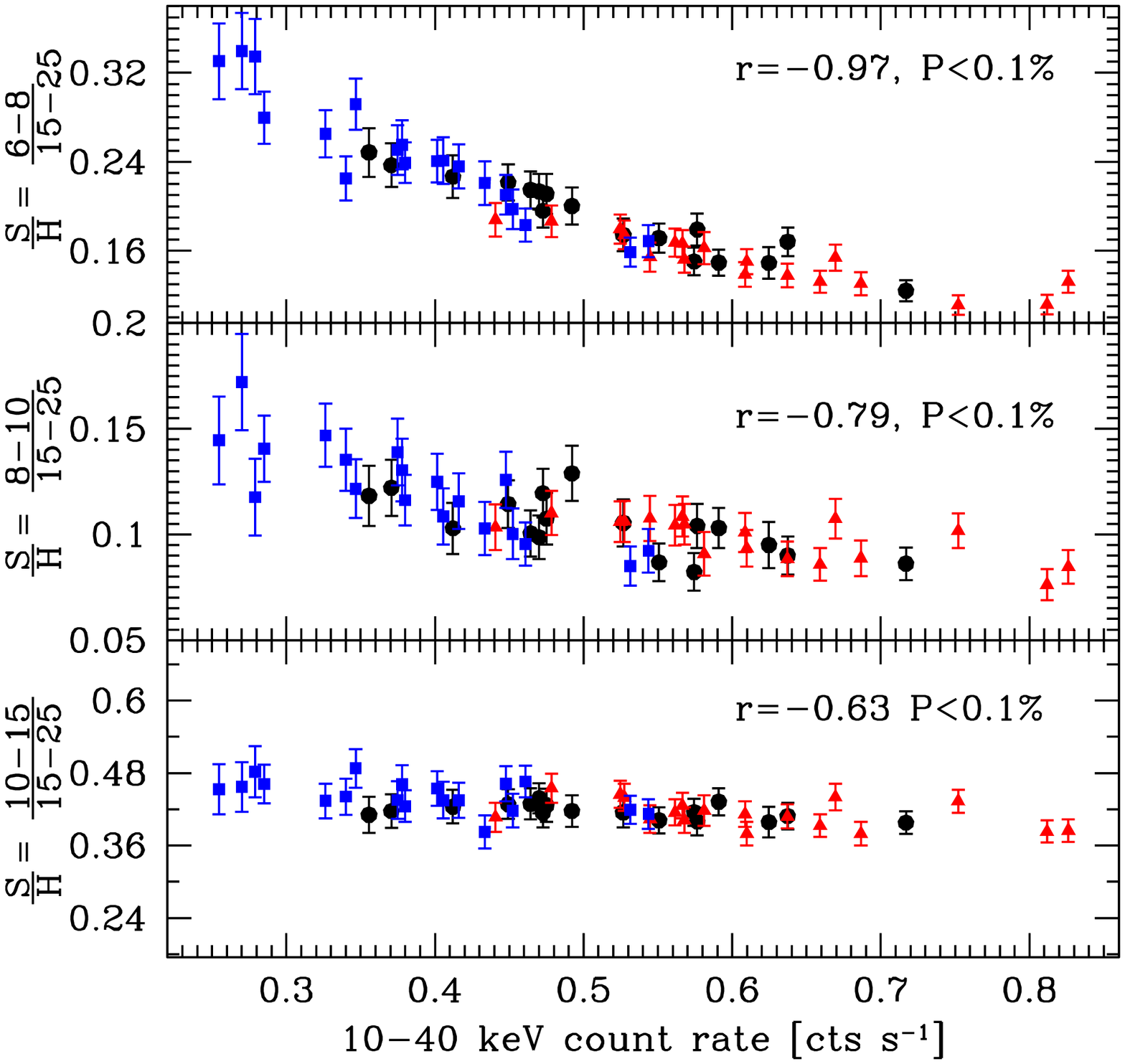}
\includegraphics[width=9.4cm]{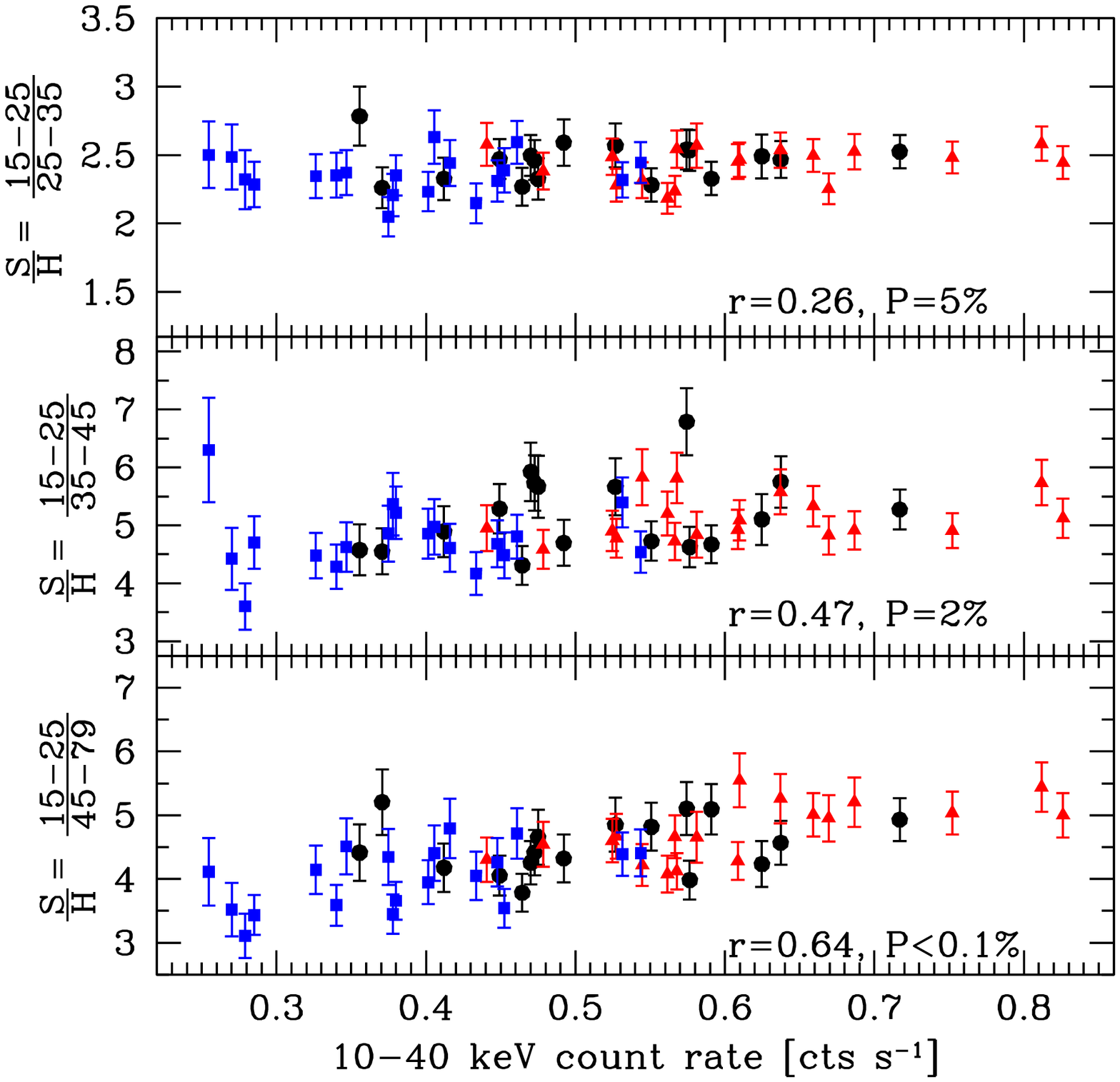}
\end{tabular}
\caption{Hardness ratio versus count rate in the $10-40$ keV range.
  Black dots, red triangles and blue squares have the same meaning as
  in Fig. \ref{lcurves}. The hardness ratio is evaluated by the relation
  ${S}\over{H}$, where S and H are the count rates in two selected energy
  ranges, H being harder than S. {\it Left panel}: {\it top} ${6-8
    }\over{15-25 }$; {\it middle} ${8-10 }\over{15-25 }$;
  {\it bottom} ${10-15 }\over{15-25 }$. {\it Right panel}: {\it
    top} ${15-25 }\over{25-35 }$; {\it middle} ${15-25
    }\over{35-45 }$; {\it bottom} ${15-25 }\over{45-79 }$.
  Each panel shows the Spearman
  rank correlation coefficient $r$ between the total hardness ratio and
  the count rate in the energy range $10-40$ keV; $P$ is the probability
  that the correlation is not statistically significant, evaluated by
  the Student's $t$ test.}
\label{hardness}
\end{figure*}

At energies below $8$~keV the light curves are approximately constant,
while above $>10$~keV the variability pattern is roughly self--similar
up to energies of the order of 45 keV.  Therefore, to investigate
possible spectral changes, the count rates in the 10--40 keV band were
considered. This energy range represents the best trade--off between
counting statistics and variability pattern.  The hardness ratios
versus the $10-40$ keV count rates are shown in
Fig. \ref{hardness}. The largest amplitude variability (a factor 3.3)
is observed in the $6-8$ keV vs. $15-25$ keV ratio. Significant
variability is also observed in the ${15-25 }\over{35-45 }$ and
${15-25 }\over{45-79 }$ ratios, while the spectral shape in the $\sim
15-35$ keV region is similar for the three observations and is
independent of the $10-40$ keV count rate.

\begin{figure}
\begin{center}
\includegraphics[width=8.5cm,height=7.5cm]{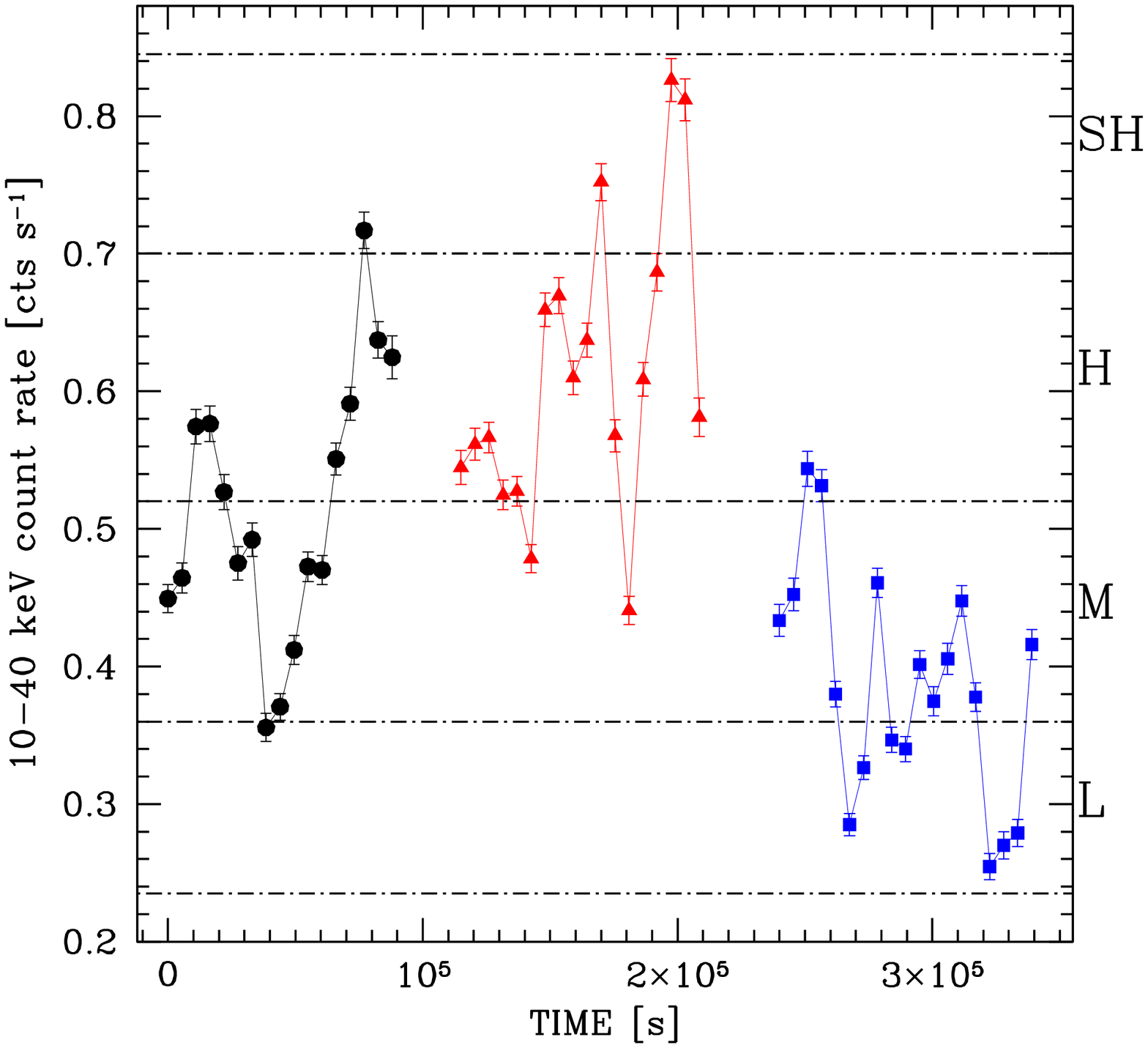}
\caption{Count--rate light curves in the $10-40$ keV energy range, in
  bins of 5500 s ( $\sim 1$ satellite orbit). The symbols are the same
  as in Fig. \ref{lcurves}. The light curves are divided into four count
  rate ranges, marked by the dot--dash lines and capital letters on
  the right (see \S~5.1).}
\label{lc1040}
\end{center}
\end{figure}

The {\it NuSTAR} $10-40$ keV count rate varies by almost a factor of 4
over the three observing periods, and by factors of up to $\sim 1.5$ on
timescales as short as $\sim 2900$ s with doubling/halving timescales
of the order of $\sim 16$ ksec (see Fig. \ref{lc1040}).

The constant count rate level of the 6--8~keV light curve, which
includes a strong iron line emission (see Yaqoob 2012 and reference
therein), suggests a constant contribution by the reflection continuum
emission, if most of the iron line emission originates from the same
cold reflector (see Marinucci et al. 2012).

The hardness ratio analysis may provide valuable hints on the spectral
components responsible for the observed variability. The 10-40 keV
count rate is strongly anticorrelated with the ${6-8 }\over{15-25 }$
ratio, and exhibits a slighter anticorrelation with the ${8-10
}\over{15-25 }$ and ${10-15 }\over{15-25 }$ ratios. This behavior and
the constant count rate level of the $6-8$~keV light curves suggest
variations of the flux level of the primary emission and/or of the
column density of the obscuring circumunclear material. On the other
hand, the correlation between the $10-40$~keV flux and the ${15-25
}\over{35-45 }$ and ${15-25 }\over{45-79 }$ ratios suggests that the
photon index of the power--law steepens while the $10-40$ keV count
rates increases.

The spectral variability is investigated in detail by means of count
rate resolved spectral analysis as described in \S~5.

\section{{\it NuSTAR} Power Density Spectrum}

We produced background--subtracted light curves in the $10-79$ keV
band for each observation, with FPMA and FPMB analyzed separately (see
\S~3). We calculated the power spectrum for each focal plane
module, combining the three observations, using the method described
in Ar{\'e}valo et al. (2012). Since each observation covers a
timescale of about $10^5$ s and gaps between the observations are
orders of magnitude longer, only frequencies above $10^{-5}$ Hz are
effectively probed.  

\begin{figure}
\centering
\includegraphics[width=7cm]{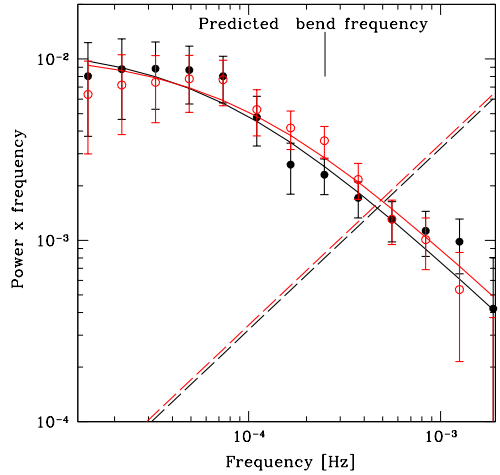}
\caption{Power density spectra in the $10-79$ keV band for FPMA (black
  filled circles) and FPMB (red open circles). The error bars
  represent only the stochastic nature of red--noise light curves. The
  solid lines represent the best--fitting models for each focal plane
  module. The dashed lines represent the Poisson noise level for each
  focal plane module, which has been subtracted from the power density
  spectra. In the plot the constant Poisson noise appears as a
  linearly increasing function and the $-1$ slope of the low--frequency
  power spectrum appears horizontal. }
\label{pds}
\end{figure}

The Poisson noise produced by the finite number of counts in each time
bin introduces additional white noise variability in the light
curves. This Poisson noise therefore adds a constant amount of power
to each frequency bin in the power spectra. We estimated the average
Poisson noise power from the high--frequency end of each power
spectrum, obtaining a value within 10\% of a simple estimate based on
the light curve errors. This average Poisson power was then subtracted
and the resulting power spectra and the Poisson noise are shown in
Fig. \ref{pds}.  Since the two focal plane modules function
simultaneously, the two power spectra represent the same intrinsic
variations in the flux and only differ due to effects of Poisson
noise. The largest differences are around timescales of a few thousand
seconds, which corresponds to the orbital gaps. These timescales are
less well sampled and more uncertain.

The power spectra are dominated by Poisson noise only at the
highest frequencies, above $10^{-3}$ Hz. Below this frequency the
power spectra rise steeply, following an approximate power--law
behavior.

We fitted each power spectrum with a bending power--law model $P$($f$)
= $A f^{\alpha_{\rm L}}$ [ 1 + ($f/f_{\rm b}$)$^{(\alpha_{\rm L} -
    \alpha_{\rm H})}$]$^{-1}$, as normally used for lower energy
X--ray power spectra of AGN (see McHardy et al. 2006 and references
therein). We obtained good fits ($\chi^2$ probability $\sim 2.5\%$ and
$\sim 10\%$, for FPMA and FPMB) fixing the low--frequency slope
$\alpha_{\rm L}$ at $-1$, high--frequency slope $\alpha_{\rm H}$ at
$-2$ and allowing the bend frequency and normalization to vary. The
best--fitting models for each focal plane module are shown in
Fig. \ref{pds}. The bend frequency 90\% error bands for two parameters
of interest are $2.8\times 10^{-5}-2.0\times 10^{-4}$ Hz for FPMA and
$4.5\times 10^{-5}-1.8\times 10^{-4}$ Hz for FPMB; as expected, the
simultaneous observations produce consistent power spectrum
parameters. These results are marginally consistent with the {\it
  RXTE} findings (Mueller et al. 2004).

In unobscured AGN, the power density spectrum bend 
frequency scales with black hole mass and accretion rate as

\begin{equation}
\log{T_B}=2.1 \log{M_{\odot,6}}-0.98\log{L_{\rm bol,44}}-2.28
\end{equation}
where $T_B$ is the break timescale in days, $M_{\odot,6}$ is the black hole mass
in units of $10^6$ M${_\odot}$ and $L_{\rm bol,44}$ is the bolometric
luminosity in units of $10^{44}$ erg s$^{-1}$ (see McHardy et
al. 2006).

Using a black hole mass of $1.4 \times 10^6 M_{\odot}$ (Greenhill et
al. 1997) and a nuclear luminosity of $L_{\rm bol} = 0.1 L_{\rm Edd}$
(Madejski et al. 2000, Guainazzi et al. 2000), relation (1) predicts a
bend timescale of $T_B=0.05$ days, corresponding to a bend frequency
of $f_B=2.3\times 10^{-4}$ Hz. This is larger than our best--fit value
in the $10-79$ keV band ($\sim 1\times 10^{-4}$ Hz). Nevertheless,
considering the bend frequency 90\% uncertainties and the intrinsic
scatter in the compilation of data that leads to the relation in
McHardy et al. (2006), we conclude that the measured power density
spectrum of \textsc{NGC4945} is in good agreement with the behavior
of the power--law continuum in unobscured sources.

An attempt to build power density spectra for the four individual
states obtained by dividing the {\it NuSTAR} light curves into four
count rate ranges (see Fig. \ref{lc1040} and \S~5) was limited due to
the short time interval of each state and the stochastic nature of the
variability. The power density spectra of each state are consistent
and no clear trend can be detected between flux level and power
density spectral shape or normalization. Therefore, the various
observations are consistent with arising from the same variability
mechanism.

\section{State--resolved spectral analysis}

\subsection{The spectra}

Given that the spectral shape depends on the $\sim 10-40$ keV count
rate, as seen in Fig. \ref{hardness}, and the $10-40$ keV count rate
varies by almost a factor 4 over the three observing periods, we
characterize the spectral variability with a proper spectral analysis
where four states were considered. The corresponding time intervals
were chosen as a trade off between a relatively high counting
statistic for the spectral analysis and a relatively short time
interval to limit rapid variability.

\begin{figure}
\includegraphics[width=9.2cm]{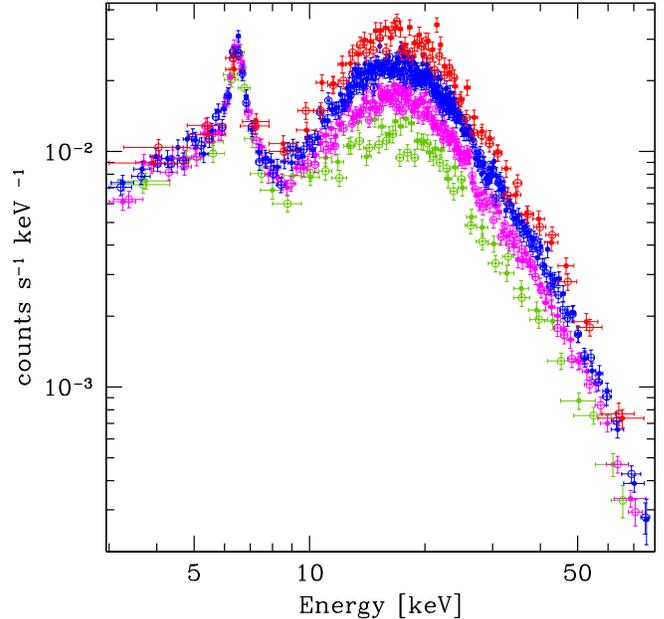}
\caption{{\it NuSTAR} spectra in the $3-79$ keV energy range. The
  solid dots are FPMA data, open dots are FPMB data. The colors
  represent: green -- L state, magenta -- M state, blue -- H
  state and red -- SH state.}
\label{spectra}
\end{figure}

We divided the observations into intervals corresponding to four count
rates in the $10-40$~keV band. As shown in Fig. \ref{lc1040}, we refer to
these count--rate intervals as the low (L), medium (M), high (H),
and super--high (SH) states. The four corresponding {\it NuSTAR}
spectra, shown in Fig. \ref{spectra}, have exposure times of $\sim 21$
ksec, $\sim 55$ ksec, $\sim 63$ ksec and $\sim 11$ ksec, respectively.

Previous X--ray observations have clearly shown that the broad ($\sim
0.5-200$ keV) spectrum of \textsc{NGC4945} is complex, requiring a
relatively large number of spectral components (see Yaqoob (2012);
  Marinucci et al. (2012) and references therein). Moreover, the
{\it NuSTAR} spectra are extracted from spatial regions which
encompass unrelated contaminating sources and extended emission.

\begin{figure}
\centering
\includegraphics[width=7cm]{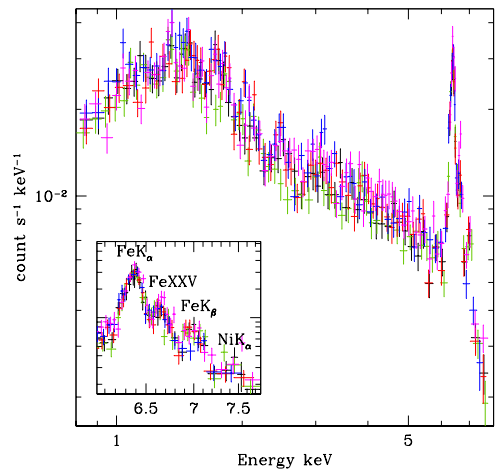}
\caption{{\it Suzaku} XIS data in the $0.8-8$ keV energy range. The
  colors represent the five observations (see
  Table~\ref{tabarchive}). The inset panel shows a zoom of the iron
  complex (i.e. Fe$K_{\alpha}$, Fe$K_{\beta}$ and Compton shoulder,
  FeXXV) and the nickel emission line.}
\label{suzaku}
\end{figure}

We made use of the excellent spatial resolution of the {\it Chandra}
observations to model the soft thermal emission due to the
starburst/X--ray plume and point--like contaminating sources, and the
good spectral resolution of {\it Suzaku} to enhance the studies of the
iron line complex at $6-7$ keV. Given that the X--ray emission of
\textsc{NGC4945} below 10 keV is relatively constant (within 10\%, see
Marinucci et al. 2012, see also Fig. \ref{suzaku}) and consistent with
the {\it NuSTAR} light curves below $\sim 8$ keV, we performed
simultaneous fits of the {\it NuSTAR}, {\it Suzaku} XIS and {\it
  Chandra} data, following the method described below (see
  Fig. \ref{speratio}). The spectral analysis for the four states (L,
M, H and SH) was performed with \textsc{xspec} (Arnaud 1996) v12.8.0.

\begin{figure}
\begin{center}
\includegraphics[width=8cm]{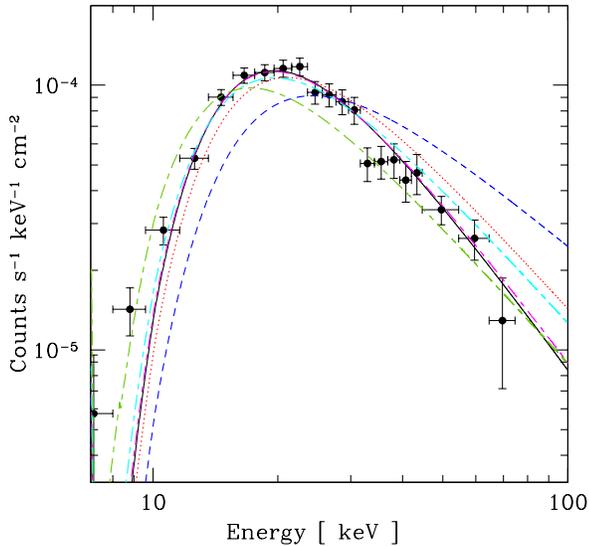}
\caption{The spectrum obtained by subtracting the L state
    spectrum from the SH state spectrum (black solid points). The
    lines show the best--fit model allowing variations of: i):
    normalization of the primary continuum (red dotted line), ii):
    photon index of the primary continuum (blue short dashed line),
    iii): absorbing column density of the primary continuum (olive
    short dashed--long dashed), iv): i)$+$ii) (black solid line), v)
    i)$+$iii) (cyan long--dashed line), vi): i)$+$ii)$+$iii) (magenta
    dotted--short dashed).}
\label{spediff}
\end{center}
\end{figure}

In order to obtain an hint of the components which dominate the
  observed spectral variability we computed the difference spectrum
  between the highest (SH) and lowest (L) count rate states.  The
  resulting spectrum (Fig.~\ref{spediff}), resembling a heavily
  absorbed power law, has statistically significant counts in the
  $\sim$8--79~keV energy range. The difference spectrum is best
  fitted with an absorbed power law (\textsc{plcabs}, Yaqoob 1997)
  $\Gamma \sim 1.9\pm 0.1$ and $N_{\rm H} \sim 4.2 \pm 0.2 \times 10
  ^{24}$ cm$^{-2}$. Given the good agreement between the best fit
  values of the difference spectrum and those obtained from the state
  resolved analysis using the same model (see \S~5.4), it is tempting to
  conclude that the observed variability pattern is mostly due to
  variations of the primary continuum.  In order to further
  investigate the origin of the variability we made the following
  test.  The SH state is fitted by varying a few parameters, whose
  initial values were determined by the L state fit . The free
  parameters are: the normalization and the photon index of the
  primary continuum, the absorbing column density and any meaningful
  combination of them.

Using a $\chi^2$ test with a minimum confidence level of 2\%, the
best--fit for the residual spectrum is obtained for variations of the
order of a  factor $\sim$ 6--8 for the normalization and $\Delta\Gamma \sim 0.2-0.3$ for the
slope of the primary continuum (see
Fig.~\ref{spediff}). The quality of the fit slightly improves ($\Delta
\chi^2 \simeq 2$ for $\Delta$d.o.f$=1$) when a variation of the column
density is included ($\Delta N_{\rm
  H}$$\sim0.1-0.2\times$ 10$^{24}$ cm$^{-2}$). 
We conclude that most of the observed variability has to be ascribed
by the intensity of the primary continuum flux and, to a lower extent,
to a steepening of the power law slope from the low to the super--high state.
If the variability were
mainly driven by absorbing column density variations, these should be
of the order of  $\Delta N_{\rm H} \sim 1.5 \times$ 10$^{24}$
cm$^{-2}$, and would have been easily recognized by this preliminary analysis.

\subsection{The model}

Thanks to the wealth of data obtained by previous observations the
main components contributing to the broad--band X--ray emission of
\textsc{NGC4945} are relatively well established. Guided by the
previous results, the broad band X--ray spectrum is modeled as
follows:

\begin{enumerate}[a)]

\item an isotropic primary continuum modelled with a power--law,
  modified by heavy photoelectric absorption and including Compton
  scattering effects (see e.g. Yaqoob 2012);

\item a cold reflection component (see e.g. Marinucci et al. 2012); 

\item the iron complex (i.e. Fe$K_{\alpha}$ at 6.4~keV, Fe$K_{\beta}$
  at 7.06~keV and the Compton shoulder) and the nickel line at
  7.47~keV detected through a simultaneous fit of the five XIS
  observations (see Fig. \ref{suzaku} {and e.g. Itoh et al. 2008});

\item a multi--temperature thermal plasma to account for the nuclear
  and extended starburst region, where the hottest component includes
  the FeXXV emission line at $\sim 6.66$~keV (Schurch et al. 2002, Done
    et al. 2003); and

\item an absorbed soft power--law plus low temperature plasma ($kT
  \sim 0.2$ keV) to account for the contaminating sources as described
  in \S~2.2.

\end{enumerate}

Preliminary fits of the five {\it Suzaku} XIS observations,
  including also the contamination component {\bf c}, indicate that
  the Fe$K_{\alpha}$, FeXXV and Fe$K_{\beta}$ emission lines are
  statistically significant at 99.99\% confidence level according to
  the F--test, while the Nickel line improves the fit at 99.95\%
  confidence level.

The ionized material from which FeXXV originates could be either
predominantly plasma collisionally heated by the starburst activity
around the nucleus (Schurch et al. 2002; Done
et al. 2003), or photoionized by the nucleus (Itoh et
al. 2008, Yaqoob et al. 2012). In either case it will contain 
free electrons that scatter a fraction of the nuclear light,
producing a continuum with approximately the same shape as the
incident AGN emission, i.e. a power law (Done et al. 2003, Bianchi et
al. 2006).

Itoh et al. (2008), Yaqoob et al. (2012), Marinucci et al. (2011) using {\it
  BeppoSAX} and/or {\it Suzaku} data, based their analysis on the
photoionization case, including in their overall model a warm
scattering power-law, which is likely to include the contribution of
serendipitous sources in the \textsc{NGC4945} field. From a spectral
point of view a warm scattered power law can be hardly distinguished
from the summed spectrum of serendipitous contaminating sources.

Done et al. (2003), thanks to the {\it Chandra} high angular
resolution, found that the starburst rather than the photoionization
scenario better describes the data, consistently with previous results
(Schurch et al.  2002). They also found that any soft scattered
component from the hot gas is negligible compared to its diffuse
emission (i.e. a factor $\sim 10^{5}$ smaller than the primary
continuum). Moreover, at low X--ray energies, the recombination lines
expected in the photoionization scenario are not detected in the
\textsc{NGC4945} reflection grating spectrometer (\textsc{RGS} onboard
       {\it XMM-Newton}) (Guainazzi \& Bianchi 2007). For these
       reasons, we believe that the soft X--ray spectrum of
       \textsc{NGC4945} is best described by thermal emission from a
       hot plasma due to starburst activity which is commonly observed
       in starburst galaxies (e.g.  \textsc{NGC253} Pietsch et
       al. 2001, \textsc{NGC6240} Boller et al. 2003). As a
       consequence, a warm scattering power-law is not considered any
       further.

All the spectral components are redshifted using $z$$=$0.001878 (see
e.g. Mouhcine et al. 2005, Mould \& Sakai 2008, Tully et al. 2008,
Jacobs et al. 2009, Tully et al. 2009, Nasonova et al. 2011).

\subsubsection{Primary continuum, cold reflection and emission lines: model components a, b, c}

Fits were performed using the \textsc{mytorus}
model\footnote{http://www.mytorus.com/} (Murphy \& Yaqoob 2009; Yaqoob
\& Murphy 2011). Spectral fits with reflection and transmission models
previously used in the literature (\textsc{pexrav}, and
\textsc{plcabs}) are briefly discussed for comparison purposes in
\S~5.4.

Assuming that the direct continuum emission is a power--law, three
tables are needed to run spectral fits: \textsc{mytorusZ} (i.e. model
component a), \textsc{mytorusS} (i.e. model component b), and
\textsc{mytorusL} (i.e. model component c). The first is a
multiplicative table that contains the pre--calculated transmission
factors that distort the direct continuum at all energies owing to
photoelectric absorption and Klein Nishina scattering (see \S~5.2 of
the \textsc{mytorus} manual); \textsc{mytorusS} and \textsc{mytorusL}
represent the scattered/reflected continuum towards the line of sight
and the emission lines (i.e. Fe$K_{\alpha}$, Fe$K_{\beta}$ and Compton
shoulder), respectively. These line tables are made with a range of
energy offsets for best--fitting the peak energies of the emission
lines. Extensive testing showed that an offset of $+$10 eV is optimal
for the \textsc{NGC4945} {\it NuSTAR}, {\it Suzaku} and {\it Chandra}
data. This covers both inter--calibration instrumental energy offsets,
and takes into account any residual offset due to a blend of neutral
and mildly ionized iron.

\textsc{mytorus} does not include a cut--off energy, but allows
different termination energies. The high-energy cut--off in
  \textsc{NGC4945} is almost completely unconstrained. Lower limits
  were reported on the basis of {\it Suzaku} ($>$ 80~keV) and {\it Swift} BAT ($>$
  100 keV) spectra by Itoh et al. (2008) and Yaqoob (2012)
  respectively. A re--analysis of  {\it BeppoSAX} data suggests that the lower 
limit on the cut--off energy could be as high as 200 keV in agreement
with previous findings (Guainazzi et al. 2000).
Therefore, in the following we used a termination energy of
500 keV, corresponding to no cut--off within the {\it NuSTAR} band.
Tests were also made with a termination energy of 200 keV without
noticeable differences.

The original geometry assumed in the Monte Carlo calculations to
generate the \textsc{mytorus} tables is that of a uniform torus with a
circular cross section; the diameter is characterized by the
equatorial column density $N_{\rm H}$, and the opening angle with
respect to the axis of the system is fixed to $60 ^{\circ}$,
corresponding to a covering factor of 0.5.

In the case of \textsc{NGC4945} the high $\tau_{\rm
    Thomson}$\footnote{$\tau_{\rm Thomson} =x \times \sigma_{\rm
      Thomson} \times N_{\rm H}$, where x is the mean number of
    electrons per H atom, which is $\sim 1.2$ assuming cosmic
    abundance.} ($2-3$) and the rapid high energy variability suggest
  that the absorber has a small effective global covering factor
  ($\sim$0.1-0.2) and a different geometrical configuration with
  respect to  the  toroidal geometry 
  (see e.g. Madejski et al. 2000, Done et al. 2003, Yaqoob et
  al. 2012). For this reason, we run \textsc{mytorus} routines in a
  ``decoupled'' mode  mimicking a clumpy 
  absorber with an arbitrary effective global covering factor (see
  Yaqoob 2012 for more details). A  few tests with \textsc{mytorus} in
  the standard configuration
 were performed for a comparison with previous results and are
 discussed in \S 5.4.

In the ``decoupled'' mode configuration the inclination angle
between the observer's line--of--sight and the symmetry axis of the
torus (hereafter $\theta_{\rm obs}$) is decoupled from the column
density $N_{\rm H}$ intercepted by the direct continuum. In this case,
the direct continuum is purely a line--of--sight quantity,
independent of geometry, and the angle $\theta_{\rm obs}$ 
is fixed to $90 ^{\circ}$. In a patchy distribution of
clouds in the background and/or from a uniform distribution with a
favourable geometry, part of the reflection from the inner far side of
the reprocessor could be unobscured by material on the near side of
the obscuring material. In this case, the far--side reflection, at least below
$\sim$10 keV, could dominate the observed spectrum. This
back--reflected continuum and the associated lines are parameterized
with a \textsc{mytorus} face--on reflection spectrum, obtained fixing
$\theta_{\rm obs}$ to $0^{\circ}$. Instead, the forward scattered
emission and associated emission lines are approximated using a
\textsc{mytorus} edge--on reflection spectrum, obtained fixing
$\theta_{\rm obs}$ to $90^{\circ}$.

In the most general case, the column density $N_{\rm H}$ obscuring the direct
continuum can be decoupled from the column density 
responsible for the back--reflection or the
forward--reflection or both.  As shown in \S 5.3, leaving the $N_{\rm
  H}$ associated to the various components free to vary does not
improve the fit. Moreover the various column densities are consistent
each other. A single $N_{\rm H}$ value is then adopted, consistently
with Yaqoob (2012).

The \textsc{xspec} format is:
\\ \\ \textsc{zpowerlw$\times$\textsc{mytorusZ}($\theta_{\rm
    obs}=90^\circ$)+\textsc{mytorusS}($\theta_{\rm obs}=90^\circ$)
  gsmooth$\times$\textsc{mytorusL}($\theta_{\rm
    obs}=90^\circ$)+\textsc{mytorusS}($\theta_{\rm obs}=0^\circ$)+
  gsmooth$\times$\textsc{mytorusL}($\theta_{\rm obs}=0^\circ$)+zgauss}
\\

The zgauss component models the Ni$K_{\alpha}$ line, which is not
included in the \textsc{mytorus} emission lines. In the following we
refer to the normalization (i.e. photons keV$^{-1}$cm$^{-2}$s$^{-1}$)
of the primary direct continuum, forward--scattered continuum,
forward--scattered emission lines, back--scattered continuum,
and back--scattered emission lines as $A_{Z90}$, $A_{S90}$,
$A_{L90}$, $A_{S0}$ and $A_{L0}$, respectively. The normalization of
the scattered (reflected) continua and the corresponding emission
lines are linked together $A_{L90}=A_{S90}$ and $A_{L0}=A_{S0}$.

\subsubsection{Multi--temperature thermal plasma: model component d}

We jointly fit the three {\it Chandra} ($0.5 - 8$~keV) datasets using
the best--fit parameters for the high-energy emission (components {\bf
  a, b, c}; \S~5.2.1) with an additional multi--temperature thermal
plasma (component {\bf d}). The photon index and the absorbing column
density $N_{\rm H}$ of the continuum are fixed to the mean values
evaluated by the spectral analysis described in the previous section
($\Gamma$=1.9, $N_{\rm H}$$=$3.55 $\times$ 10$^{24}$ cm$^{-2}$).

Following Schurch et al. (2002), the soft thermal emission due to the
starburst/X--ray plume is modelled with a three--temperature,
optically thin thermal plasma, allowing the absorption of each
component to vary independently. The \textsc{xspec} format for
component {\bf d} is:
\\ \\ \textsc{\textsc{constant}$\times$\textsc{wabs\_1}$\times$(\textsc{mekal\_1}
  + \textsc{wabs\_2}$\times$\textsc{mekal\_2} +
  \textsc{wabs\_3}$\times$\textsc{mekal\_3})} \\ \\ where the
\textsc{wabs\_1} component is fixed to Galactic column density and the
constant is the normalization factor between the {\it Chandra}
observation performed in 2000 and the two performed in 2004. The
normalization factor is 1.00$\pm_{0.05}^{0.07}$, confirming that the
\textsc{NGC4945} X--ray emission below 10 keV is relatively
constant. In the following the normalization is fixed to unity.

\begin{figure}
\centering \includegraphics[width=9cm]{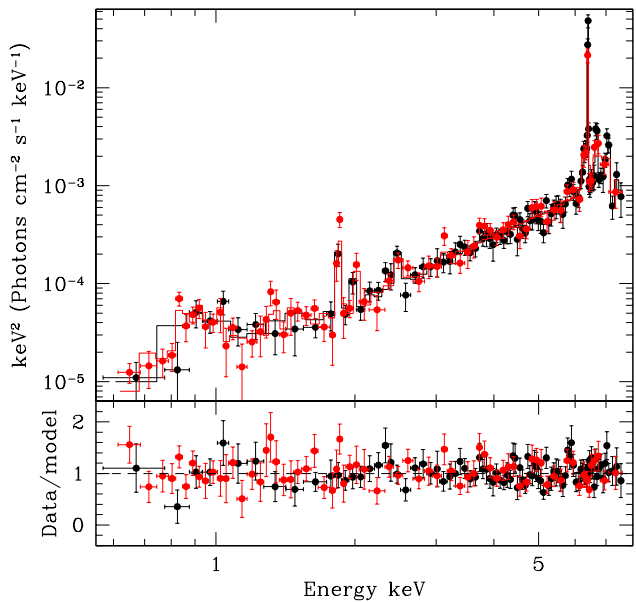}
\caption{The {\it Chandra} spectra, unfolded with the instrument
  response, of the \textsc{NGC4945} nucleus plus X--ray plume; with
  the two co--added 2004 observations performed by HETG (zeroth) shown
  in black, and the 2000 observation performed by ACIS--S shown in red
  (see Table~\ref{tabarchive}). Lower panel shows the data/model
  residuals.}
\label{spechandra}
\end{figure}

Our {\it Chandra} best--fit model (Fig. \ref{spechandra} and
Table~\ref{tabchandra}) is fully consistent with previous results
(Schurch et al. 2002, Done et al. 2003). Model component {\bf d} in
the final spectral analysis for the four states (L, M, H and SH) is
fixed to the {\it Chandra} best--fit parameters.

\begin{table}
\footnotesize
\caption{Multi--temperature thermal plasma: {\it Chandra} best--fitting parameters}
\begin{center}
\begin{tabular}{lc}
\hline
Parameter  & value \\ 
\hline
$N_{\rm H}$$_1$$^a$ & 0.157\\
$kT_1$$^b$&       0.67$\pm_{0.07}^{0.11}$ \\                 
$K_1$$^c$ &        2.1$\pm_{0.4}^{0.3}$ $\times$ 10$^{-5}$\\  
$N_{\rm H}$$_2$$^a$ &   2.0$\pm_{0.4}^{0.6}$  \\                   
$kT_2$$^b$&       0.9$\pm_{0.2}^{0.2}$  \\                   
$K_2$$^c$&        5.3$\pm_{1.6}^{3.5}$ $\times$ 10$^{-4}$\\  
$N_{\rm H}$$_3$$^a$ &   10.5$\pm_{3.3}^{4.9}$\\                    
$kT_3$$^b$&       4.0$\pm_{1.6}^{2}$\\                       
K$_3$$^c$&        8$\pm_{2}^{8}$ $\times$ 10$^{-4}$\\        
PHA bins   & 171 \\
degrees of freedom     & 160  \\
$\chi^2$      &  137.2 \\
\hline
\end{tabular}
\end{center}

The table reports the {\it Chandra} best--fit values with uncertainties
at the 90\% confidence level for one parameter of interest ($\Delta
\chi^2$=2.706) for the model component {\bf d}.

$^a$ Column density $N_{\rm H}$ in units of 10$^{22}$ cm$^{-2}$. The first
component, \textsc{wabs\_1}, is fixed to Galactic column density
(Heiles \& Cleary 1979). $^b$ Temperature in keV; $^c$ normalization at 1 keV in
unit of photons keV$^{-1}$ cm$^{-2}$s$^{-1}$ for the spectral
components. Subscripts refer to the spectral component number (see
\S~5.2.2).

\label{tabchandra}
\end{table}

\subsubsection{Contaminating sources: model component e}

The effect of contamination due to individual sources within the
{\it NuSTAR} and {\it Suzaku} extraction regions is parameterized with an
absorbed power--law and low temperature plasma plus absorption
(see \S~2.2). To this aim, the {\it NuSTAR} ($3-79$ keV), {\it Suzaku}
XIS ($0.8-7.8$ keV) and {\it Chandra} ACIS ($0.5-7.8$ keV) data are
jointly fitted using the spectral components {\bf a, b, c} and {\bf d},
while component {\bf e} is applied only to the {\it NuSTAR} and {\it Suzaku}
data.

The mean values of the best--fitting parameters of the absorbed
power--law are: $\Gamma=2.22\pm0.01$, $N_{\rm H}=(1.26\pm0.14)$
$\times$ 10$^{22}$ cm$^{-2}$ and a normalization at 1 keV of
$(8.2\pm0.9) \times 10^{-4}$ photons keV$^{-1}$ cm$^{-2}$s$^{-1}$. The
corresponding values for the thermal component are: $kT=0.17\pm0.01$
keV, $N_{\rm H}=(0.81\pm0.12) \times 10^{22}$ cm$^{-2}$ and
normalization $(2.0\pm0.6) \times 10^{-2}$ photons keV$^{-1}$
cm$^{-2}$s$^{-1}$. The {\it NuSTAR} $4-10$ keV flux of the component
{\bf e} is $\sim 10^{-12}$ erg s$^{-1}$ cm$^{-2}$. The {\it Suzaku}
contamination flux is within $\pm$10\% of the {\it NuSTAR} one. In the
final spectral analysis for the four states (L, M, H and SH), the
parameters of spectral component {\bf e} are fixed to the best--fit
values reported above.

Considering all the model components, the best--fit values of the
normalization factors for the {\it Suzaku} observations with respect
to the first observation are 1.054$\pm_{0.07}^{0.08}$,
0.984$\pm_{0.08}^{0.08}$, 1.037$\pm_{0.08}^{0.08}$,
1.01$\pm_{0.07}^{0.07}$, respectively. The normalization of the first
{\it Suzaku} observation with respect to the 2004 {\it Chandra}
observation is 0.97. The normalization of the 2004 {\it Chandra}
observation with respect to the {\it NuSTAR} observations is
1.07$\pm$0.05. In the final spectral analysis we only allow variations
of the normalization factor of the 2004 {\it Chandra} observation with
respect to the {\it NuSTAR} spectra; we fix the normalizations of the
2000 {\it Chandra} and {\it Suzaku} observations to the ratio between
the best--fit values reported above and the 2004 {\it Chandra}
observation one.

\subsection{Results}

\begin{figure*}
\centering
\includegraphics[width=8.cm]{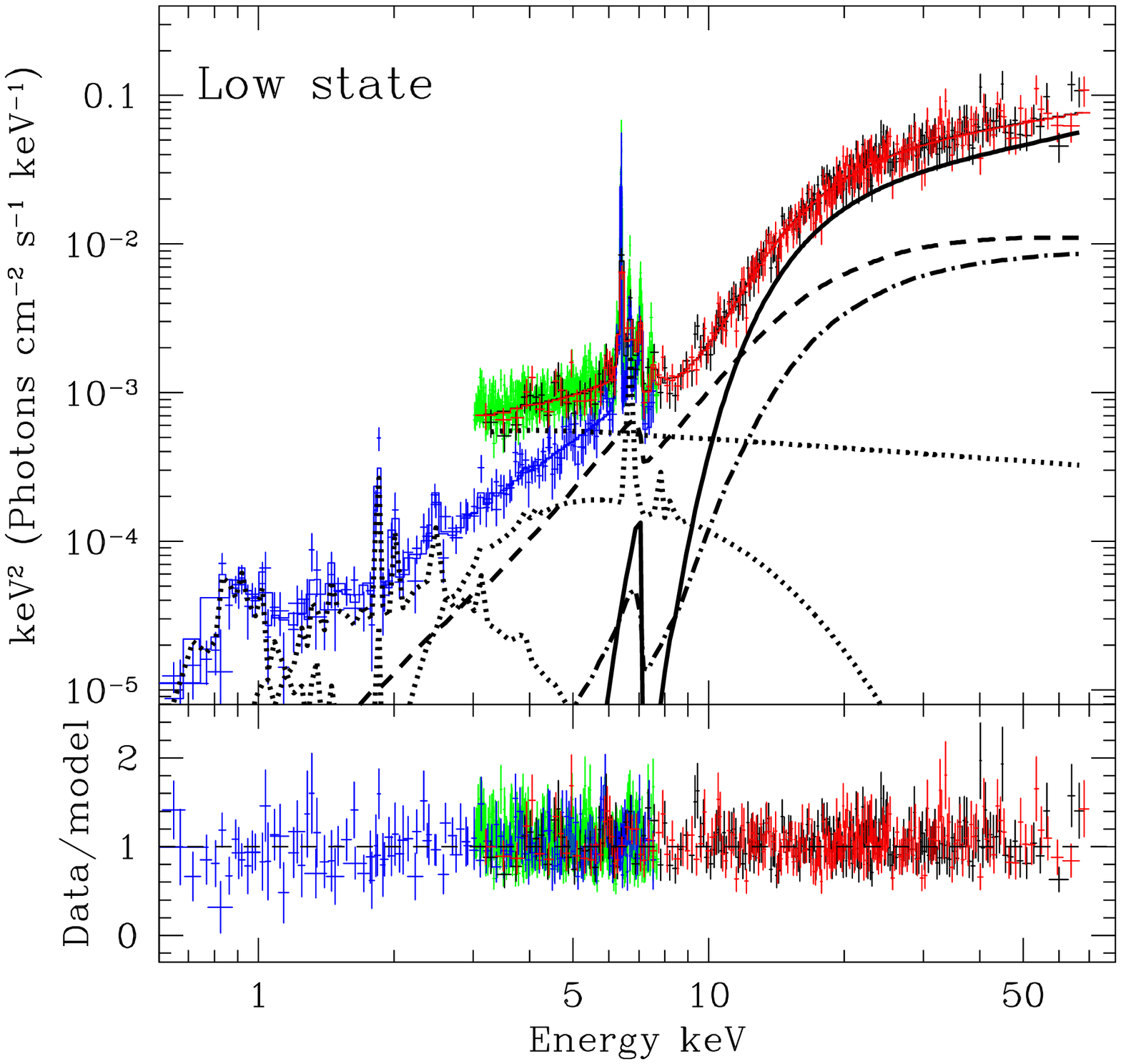}\includegraphics[width=8.cm]{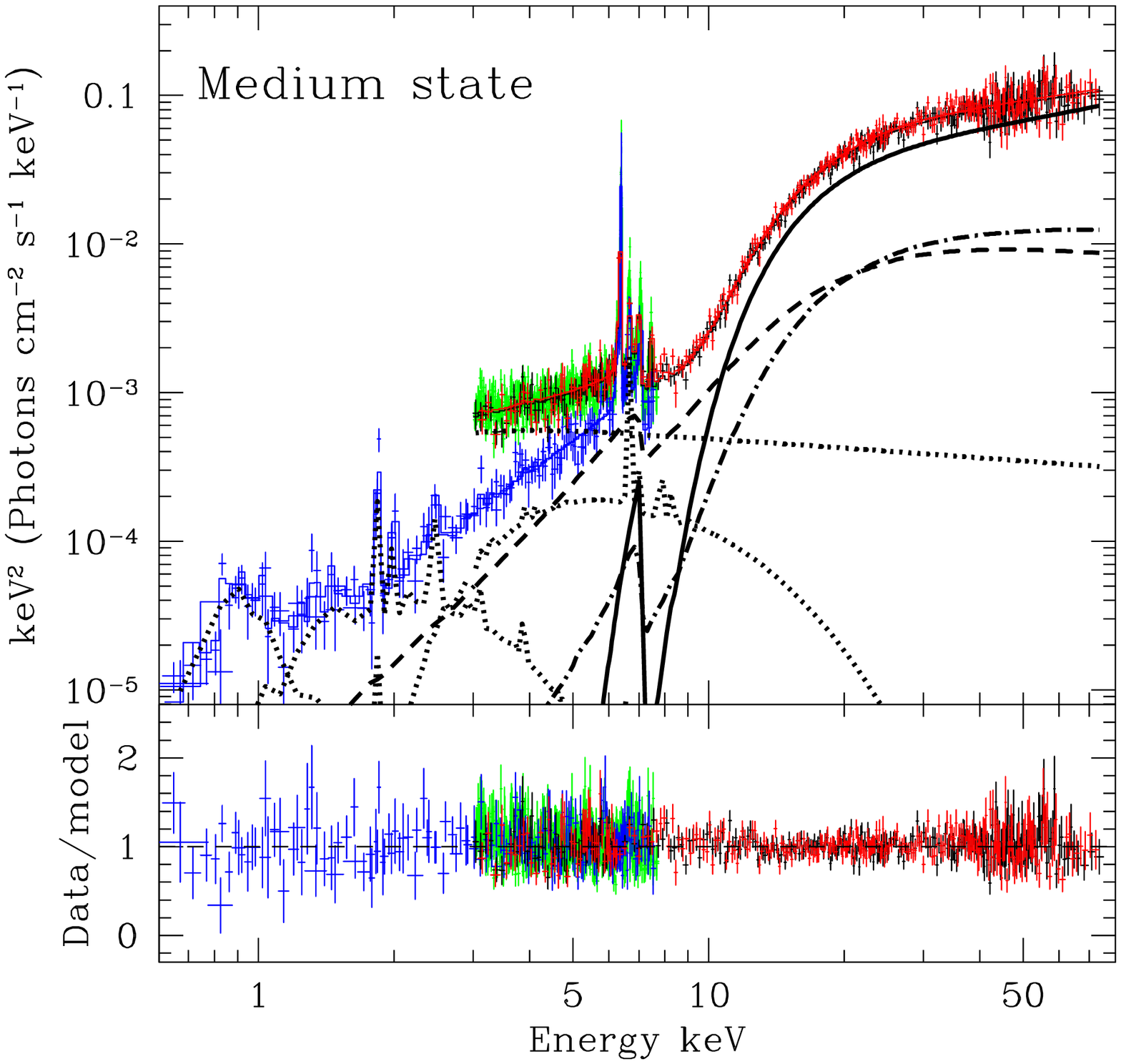}

\includegraphics[width=8.cm]{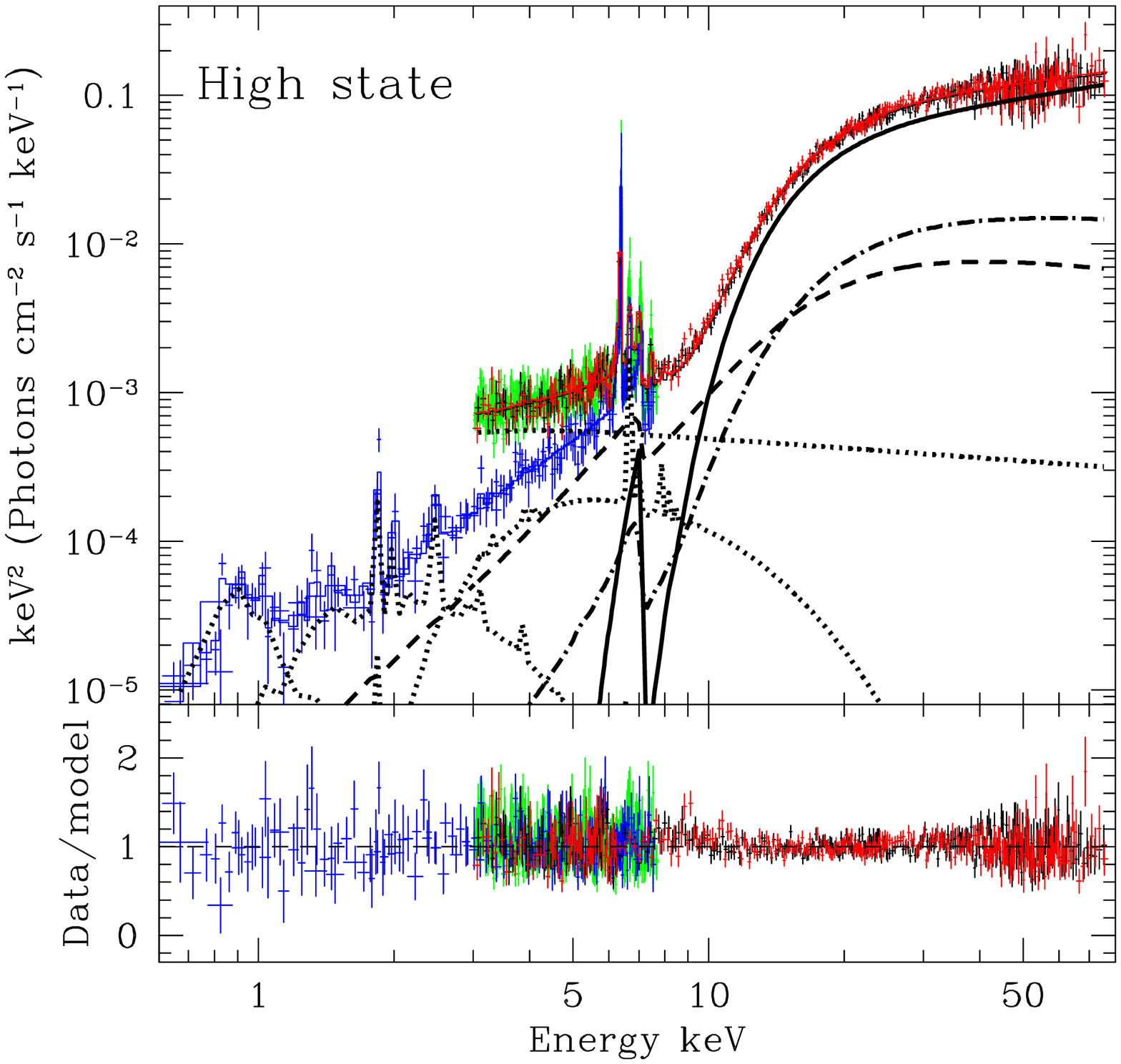}\includegraphics[width=8.cm]{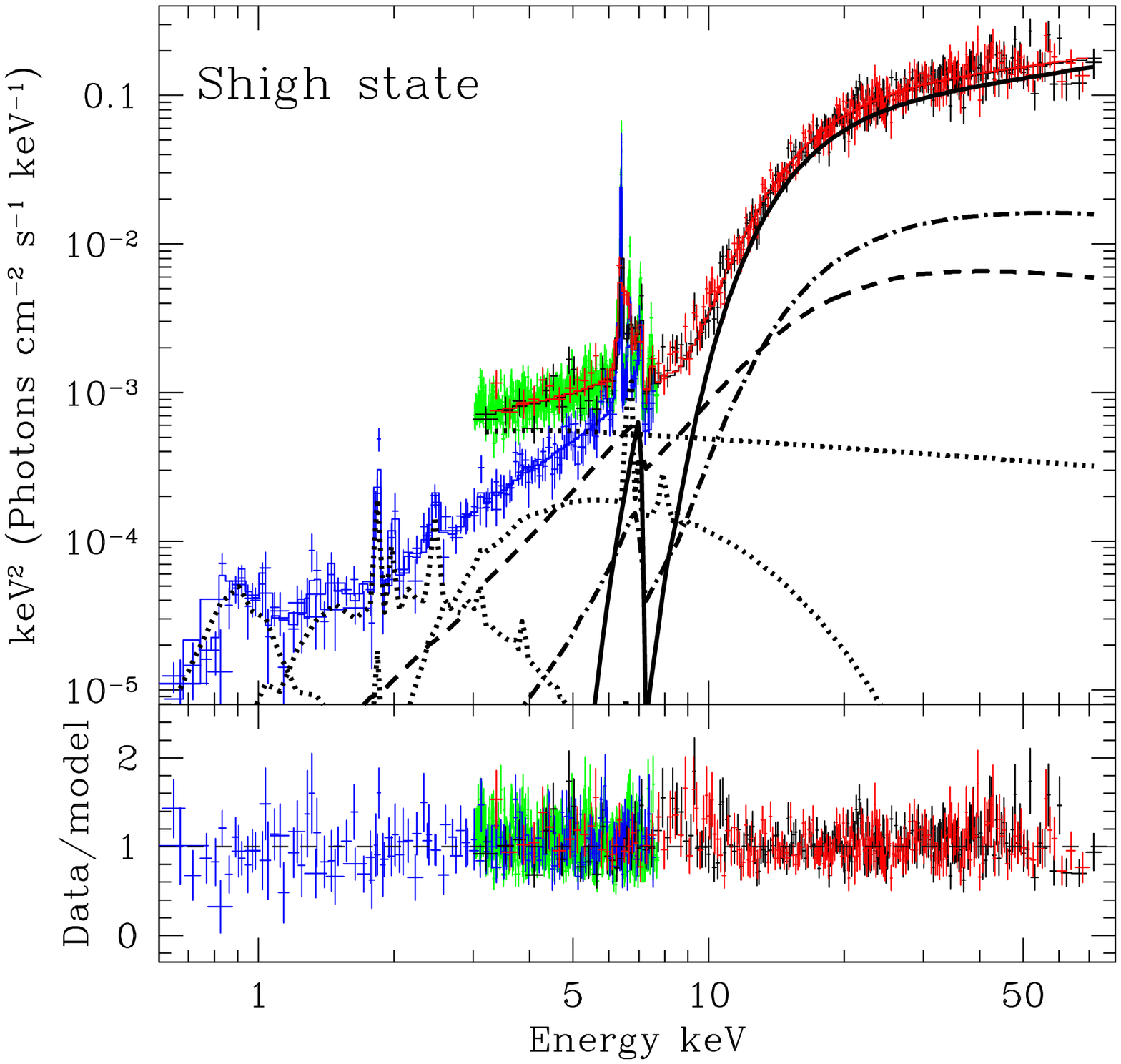}
\caption{The broad--band best--fit spectra, models (upper
    panels) and residuals (lower panels) for the L, M, H and SH states
    (from top to bottom, clockwise). The {\it NuSTAR} (black is FPMA,
    red is FPMB), the five {\it Suzaku} (green) and {\it Chandra}
    (blue) spectra unfolded with the instrument responses (for
    single {\it Suzaku} and {\it Chandra} spectra see
    Fig. \ref{suzaku} and \ref{spechandra}, respectively) are reported
    in the upper panels. The total model, including the
    Fe$K_{\alpha}$, Fe$K_{\beta}$ and Nickel lines (not shown as
    single components for clarity), is superimposed.
    The model single continuum components are also shown: the
    primary continuum (black solid lines), the back--reflection
     (black dashed lines), the forward--reflection 
    (black dashed--dotted lines), the three thermal spectra
     and the power--law due to the contribution of the
    serendipitous sources in the \textsc{NGC4945} field (black dotted
    lines). In the lower panels the data/best--fit model ratios are
    shown.}
\label{speratio}
\end{figure*}

\begin{figure}
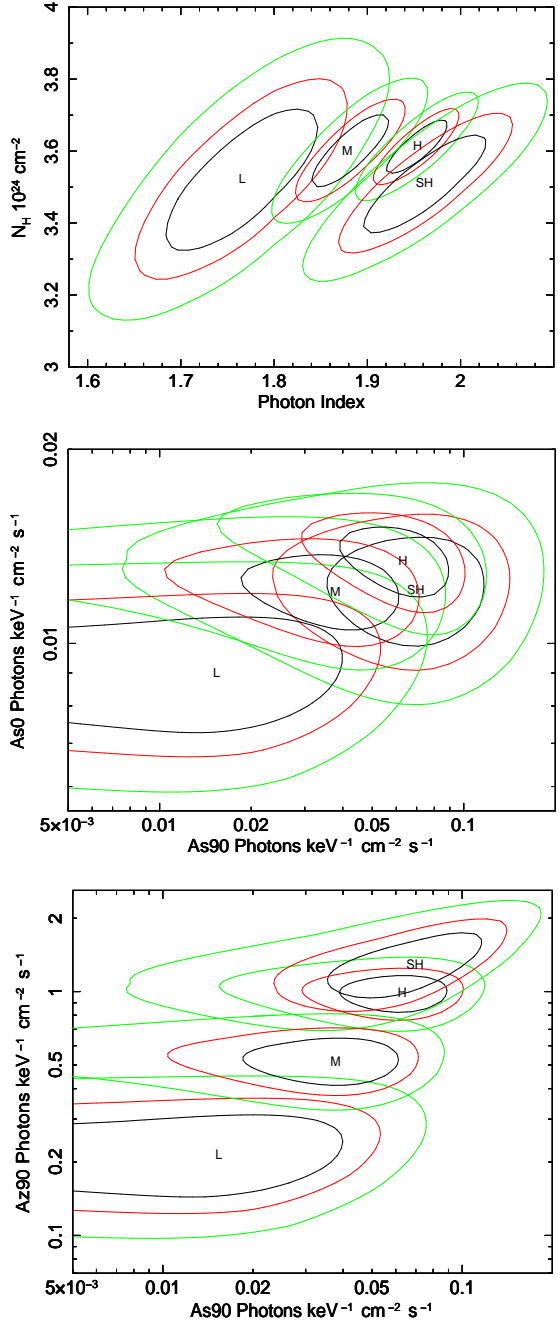

\centering
\includegraphics[width=5.5cm,angle=-90]{18.ps}
\vskip+5pt
\includegraphics[width=5.7cm,angle=-90]{19.ps}
\vskip+13pt
\hskip+13pt
\includegraphics[width=5.4cm,height=7.15cm,angle=-90]{20.ps}
\vskip+15pt
\caption{68\%, 90\% and 99\% confidence contours for the L, M, H and
  SH states for different model parameters. {\it Top panel}: column
  density $N_{\rm H}$ versus photon index for the direct continuum;
  {\it middle panel}: normalization of the back--reflected continuum
  versus normalization of the forward--scattered continuum; {\it
    bottom panel}: normalization of the back--reflected continuum
  versus the normalization of the direct continuum.}
\label{modC}
\end{figure}

The spectral analysis for the four states (L, M, H and SH) is
performed jointly fitting the {\it NuSTAR} ($3-79$ keV), {\it Suzaku}
XIS ($3-7.8$ keV) and {\it Chandra} ($0.5-7.8$ keV) data, using the
five--component spectral model described above. The best--fit
parameters are reported in Table~\ref{tabfit}, the broad--band
best--fit spectra, residuals and models in the four states are
reported in Fig. \ref{speratio} and the confidence contours of the
joint errors for different parameters of interest are shown in
Fig. \ref{modC}. The nucleus is obscured by Compton--thick matter with
a remarkably constant column density $N_{\rm H}$ from the L to the SH
state of the order of 3.55 $\times 10^{24}$ cm$^{-2}$, corresponding
to $\tau_{\rm Thomson}\sim 2.9$. The primary continuum spectral slope
slightly steepens ($\Delta\Gamma \sim 0.2$) when the source gets
brighter (Fig. \ref{modC}, top panel), with a behavior typical f
Seyfert galaxies on longer timescales (e.g. Sobolewska \& Papadakis
2009, Caballero--Garcia et al. 2012). The primary continuum flux,
without considering the contribution of the constant reflected
components, in the $10-40$ keV energy range increases from the L to
the SH state by a factor of $\sim 3.3$. The flux impinging on the
Compton--thick obscuring gas and scattered into the line of sight
($A_{S90}$ in Table~\ref{tabfit}) is correlated with the primary flux,
albeit with a much lower variability amplitude (less than a factor of
1.2; see Fig. \ref{modC}, lower panel) as expected due to the smearing
effect introduced by scattering in the obscuring gas. The
forward--scattered flux becomes progressively less constrained when
the intrinsic flux becomes fainter. In the L state it is not formally
detected, though the 90\% upper limit is consistent with the values in
the M, H and SH states.

The back--reflection component ($A_{S0}$) is almost constant (within a
few 10\%) across the various states. There is no clear evidence of a
difference in the column density $N_{\rm H}$ between the matter
responsible for absorption, scattering and reflection.

Following Yaqoob (2012), in a time--steady situation, we can
  estimate a covering factor from $A_{S90}$/$A_{Z90}$ if
  $(A_{S90}$/$A_{Z90}) <<1$, as we indeed find. The original geometry assumed
  in the Monte Carlo calculations to generate the \textsc{mytorus}
  tables (i.e. $A_{S90}$/$A_{Z90} = 1$) corresponds to a covering
  factor of 0.5. Therefore, for a different configuration, the
  covering factor can be obtained by the ratio above rescaled by a
  factor 0.5. The best--fit values reported 
in Table~6 indicate a relatively constant covering factor of $\sim
0.13$.

The observed iron and nickel line intensities are consistent, but much
better constrained than previous measurements reported in the
literature. There is no evidence of variability in the line equivalent
widths and fluxes (evaluated following the method in \S~7.3.5.4
  of the \textsc{mytorus} manual). The mean values of the line fluxes
in units of 10$^{-14}$ erg cm$^{-2}$ s$^{-1}$ are $31.3\pm0.5$,
$3.2\pm0.3$, $4.5\pm0.5$ and $2.5\pm0.9$ for the Fe$K_{\alpha}$,
FeXXV, Fe$K_{\beta}$ and Nickel line, respectively. The highly ionized
helium--like iron line is also remarkably constant, suggesting it may
be associated with the hottest component of the multi--temperature
plasma.

\begin{table}
\footnotesize
\caption{Best--fitting parameters}
\begin{center}
\begin{tabular}{p{2cm}p{1.32cm}p{1.32cm}p{1.32cm}p{1.32cm}}
\hline
Parameter& L & M & H & SH \\
\hline
$\Gamma$$^a$&   1.77$\pm_{0.09}^{0.09}$&	1.88$\pm_{0.04}^{0.05}$&  1.95$\pm_{0.04}^{0.03}$&  1.96$\pm_{0.07}^{0.07}$ \\ 
$N_{\rm H}$$^b$ & 3.5$\pm_{0.2}^{0.2}$& 3.6$\pm_{0.1}^{0.1}$&  3.6$\pm_{0.1}^{0.1}$&      3.5$\pm_{0.1}^{0.1}$ \\  
 $A_{\rm Z90}$$^c$&  0.21$\pm_{0.07}^{0.11}$ & 0.51$\pm_{0.11}^{0.15}$&  0.98$\pm_{0.17}^{0.20}$&	1.28$\pm_{0.36}^{0.51}$ \\ 
 $A_{\rm S90}$$^d$&   $\le$0.04&	   0.04$\pm_{0.02}^{0.03}$&    0.06$\pm_{0.03}^{0.03}$&      0.07$\pm_{0.04}^{0.05}$ \\  
 $A_{\rm S0}$$^e$&  0.009$\pm_{0.002}^{0.002}$&	   0.012$\pm_{0.002}^{0.002}$&  0.013$\pm_{0.002}^{0.002}$&    0.012$\pm_{0.003}^{0.003}$ \\  
Nickel energy$^f$ & 7.48$\pm_{0.05}^{0.05}$& 7.48$\pm_{0.04}^{0.04}$&    7.49$\pm_{0.05}^{0.05}$&      7.48$\pm_{0.06}^{0.05}$ \\
Fe$K_{\alpha}$ EW$^g$ &  0.97$\pm_{0.29}^{0.35}$  &   0.92$\pm_{0.20}^{0.25}$  &    0.94$\pm_{0.24}^{0.17}$  &  0.94$\pm_{0.32}^{0.34}$ \\  
FeXXV  EW$^g$ &  0.10 $\pm_{0.02}^{0.10}$ &   0.08$\pm_{0.02}^{0.09}$  &    0.09$\pm_{0.02 }^{0.09}$ &  0.09$\pm_{0.02}^{0.09}$ \\  
Fe$K_{\beta}$ EW$^g$ & 0.14 $\pm_{0.04}^{0.05}$ &   0.12$\pm_{0.03}^{0.04}$  &    0.15 $\pm_{0.04}^{0.03}$ &  0.15$\pm_{0.05}^{0.05}$ \\  
Nickel  EW$^g$ & 0.11  $\pm_{0.04}^{0.05}$ &  0.13$\pm_{0.04}^{0.04}$  &    0.11$\pm_{0.04}^{0.04}$  &  0.10$\pm_{0.04}^{0.05}$ \\  
FPMB/FPMA$^h$ & 0.99$\pm_{0.03}^{0.03}$& 1.03$\pm_{0.02}^{0.02}$&   1.03$\pm_{0.01}^{0.01}$&     1.02$\pm_{0.03}^{0.03}$  \\
PHA bins   & 1054 & 1206 & 1251 & 1064 \\
d.o.f.     & 1045 & 1197 & 1242 &1055 \\
$\chi^2$      &  1070.9 & 1265.6 & 1343.8 & 1117.603\\
reduced $\chi^2$        &1.02	& 1.06	     & 1.08		    &	  1.06           		   \\  	           
$\chi^2$  probability     &28.2\%	& 8.2\%	     & 2.3\%	    &		  8.2\%	           \\ 	           
5-10 keV flux$^i$ &  1.87$\pm_{0.05}^{0.05}$ & 2.00$\pm_{0.03}^{0.03}$  & 2.05$\pm_{0.03}^{0.03}$&   2.08$\pm_{0.05}^{0.08}$\\
10-40 keV flux$^i$ & 61$\pm_{4}^{4}$ &  86$\pm_{6}^{5}$ & 121$\pm_{6}^{6}$ & 154$\pm_{11}^{9}$\\
40-79 keV flux$^i$  & 76$\pm_{11}^{4}$ & 102$\pm_{2}^{4}$ & 135$\pm_{4}^{2}$&  175$\pm_{19}^{5}$\\
L(2-10 keV)$^l$ & 12.3 & 24.8 & 42.1 &54.4\\
$\lambda_{\rm Edd}$ & 0.07 & 0.14 & 0.24 & 0.32 \\
covering factor &  0.14$\pm_{0.14}^{0.26}$ &   0.16$\pm_{0.04}^{0.06}$&       0.12$\pm_{0.03}^{0.03}$&       0.11$\pm_{0.04}^{0.04}$ \\  
\hline
\end{tabular}
\end{center}

Best--fit values with uncertainties at the 90\%
confidence level for one parameter of interest ($\Delta \chi^2$=2.706)
for states L, M, H and SH.

$^a$ Direct power--law photon index; $^b$ column density $N_{\rm H}$
in units of 10$^{24}$ cm$^{-2}$; $^c$ normalization at 1 keV of the
direct power--law in units of photons keV$^{-1}$ cm$^{-2}$s$^{-1}$;
$^d$ normalization at 1 keV of the forward--reflection component in
units of photons keV$^{-1}$ cm$^{-2}$s$^{-1}$; $^e$ normalization at 1
keV of the back--reflection component in units of photons keV$^{-1}$
cm$^{-2}$s$^{-1}$; $^f$ Nickel line energy in units of keV; $^g$ line
equivalent width in units of keV; $^h$ normalization factor between
FPMB and FPMA; $^i$ FPMA {\it NuSTAR} observed flux in units of
10$^{-12}$ erg cm$^{-2}$ s$^{-1}$; $^l$ FPMA {\it NuSTAR} intrinsic
$2-10$ keV luminosity in units of 10$^{41}$ erg s$^{-1}$ for $D=$3.8
Mpc (Mouhcine et al. 2005, Mould \& Sakai 2008, Tully et al. 2008,
Jacobs et al. 2009, Tully et al. 2009, Nasonova et al. 2011).

\label{tabfit}
\end{table}

\subsection{Comparison with standard reflection models}

For the sake of completeness, and to facilitate comparison with
previous results in the literature, we briefly describe the results
obtained by fitting a standard torus model and use simpler
approximations for the absorption and reflection in Compton--thick
media.

Fitting the \textsc{mytorus} model in the original geometry
(i.e. ``coupled'' mode, which represents a uniform torus with
half--opening angle of $60^{\circ}$ and a 0.5 covering factor) yields
significantly worse fits (total $\Delta \chi^2$ $\sim 800$ with total
degrees of freedom decreased by 4) with respect to the fits obtained
in the ``decoupled'' mode (see \S~5.3). We can therefore reject this
uniform torus model. Leaving the normalization of the
forward--scattered component ($A_{S}$) free to vary, we obtain
statistically good fits. We find that $A_{S}$ increases from
  $A_{S}\sim 0.12 \times A_{Z}$ for the SH state to $A_{S}\sim 0.5
  \times A_{Z}$ for the L state; this indicates that the
  Compton--scattered continuum is much weaker in the H and SH
  states (as also found by Yaqoob 2012 for the {\it BeppoSAX} and
  {\it Swift} BAT data). These findings provide further evidence
  against a uniform torus model.
Moreover, the L and M states
would be reflection--dominated. We can therefore reject this model as
inconsistent with the observed variability.

We also considered simpler, but unphysical approximations for the
absorption and reflection in Compton--thick media. We modelled the
primary continuum transmitted through a high column density absorber
with \textsc{plcabs} (Yaqoob 1997). We modelled the reflection
component with an exponentially cut--off power--law spectrum reflected
from neutral material with infinite column density using
\textsc{pexrav} (Magdziarz \& Zdziarski 1995). The absorption column
density $N_{\rm H}$ is slightly higher ($\sim 4 \times 10 ^{24}$
cm$^{-2}$) and fully consistent with literature fits, while the direct
continuum is slightly harder ($\Delta \Gamma \sim 0.1$). The observed
variability is due to an increase in the direct continuum emission
from the L to SH state, similar to the ``decoupled'' \textsc{mytorus}
model.

The statistical quality of the best--fit
  \textsc{plcabs}/\textsc{pexrav} model is equivalent to the best--fit
  ``decoupled'' model described earlier, being the
  \textsc{plcabs}/\textsc{pexrav} model the most similar to the
  ``decoupled'' \textsc{mytorus} model as underlined by Yaqoob (2012).
  A disc/slab geometry, with an infinite column density for the
  material responsible for the Compton--scattered continuum is assumed
  in the \textsc{pexrav} model.  The spectral features associated to
  scattering in a finite column density medium cannot be reproduced.
  Moreover, the column density of \textsc{NGC4945} is close to the
  limit of validity of the \textsc{plcabs} routines (i.e. $N_H =5
  \times 10^{24}$ cm$^{-2}$). Finally, as pointed out by Murphy \&
  Yaqoob (2009), using \textsc{pexrav} and \textsc{plcabs} may produce
  a bias towards solutions dominated by the direct continuum. For all
  these reasons, \textsc{mytorus} returns a more physical description
  of the observations (Yaqoob \& Murphy 2009, 2011, Yaqoob 2012).

\subsection{$N_{\rm H}$ variability ?}

\begin{figure}
\centering
\includegraphics[width=5.5cm,angle=-90]{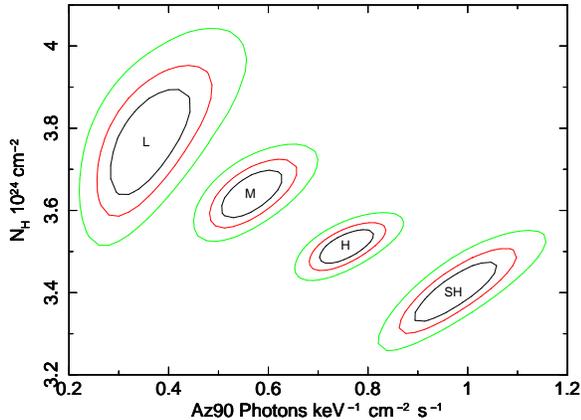}
\caption{68\%, 90\% and 99\% confidence contours for the column
    density $N_{\rm H}$ versus the normalization of the direct
    continuum, for the L, M, H and SH states. The confidence contours
    are for the ``decoupled'' \textsc{mytorus} model with the photon
    index for the direct continuum fixed to 1.9.}
\label{nhAzgfixed}
\end{figure}

The spectral analysis performed with the ``decoupled''
\textsc{mytorus} model suggests changes of the photon index for the
primary continuum, and constant column density for the obscuring
matter. This behavior could be equivalent to a constant photon index
and variable column density. Therefore, to further investigate
possible variations of obscuring matter, we performed spectral
analysis of the four states L, M, H and SH using the ``decoupled''
\textsc{mytorus} model with the photon index of the primary continuum
fixed to the mean value of $\Gamma$=1.9.  The global quality of the
fit worsened (total $\Delta \chi^2$ $\sim 18$ with total degrees of
freedom decreased by 4). The best--fit values of the column density of
the primary continuum are consistent with being constant ($ N_{\rm
  H}=3.8\pm 0.2$, $3.6\pm0.1$, $3.5\pm0.1$ and $3.4\pm0.1$ $\times$
10$^{24}$ cm$^{-2}$, for the L, M, H, and SH states, respectively),
but show a weak anti--correlation with the primary continuum (see
Fig. \ref{nhAzgfixed}). This behavior suggests that possible
variations of the column density could be due to small variations in
the ionization state of the absorber, following variations in the
intensity of ionizing radiation, rather than to variations in the
amount of absorbing gas along the line of sight. We remark that the
small variation of the absorption column density is obtained under the
hypothesis of a constant photon index and thus should be considered
with caution.

\section{Discussion}

The state--resolved spectral analysis of the three {\it NuSTAR}
observations of the bright Seyfert 2 galaxy \textsc{NGC4945} confirms
and extends the results obtained by previous observations in the hard
X--ray band. The broad--band $3-79$ keV emission is characterized by
two main spectral components. The first is primary continuum emission
piercing through a Compton--thick obscuring medium which shows
rapid variability, and which dominates the E$>10$~keV spectrum. The second
component which dominates at E$<10$~keV results from a combination of different components, all of them
relatively constant: i) a low--energy tail of the back--reflected
nuclear spectrum with neutral emission lines, ii) a
multi-temperature thermal plasma due to the nuclear starburst and
extended super wind, and iii) a summed contribution of individual
point--like sources within 75$\asec$ of the \textsc{NGC4945} nucleus,
modelled with a power--law and low temperature plasma. 

\begin{figure}
\centering
\includegraphics[width=8cm]{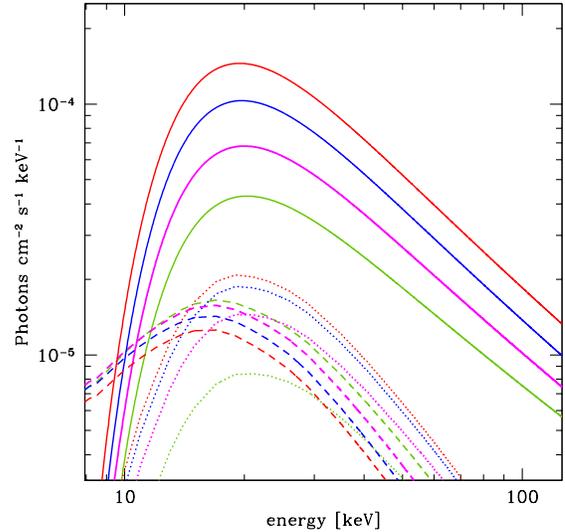}
\caption{Best--fitting model for the main hard X--ray components: the
  solid lines represent the direct continuum component, the dotted
  lines represent the forward--scattered component, and the dashed
  lines represent the back--reflected continuum component. As in
  Fig.~\ref{spectra}, the green, magenta, blue and red lines represent
  the L, M, H and SH states, respectively.}
\label{mod}
\end{figure}

Given the strong dependence of the spectral shape on the $10-40$ keV
flux, we extracted four states based on the brightness, which we call
L, S, H and SH (see Fig. \ref{lc1040}), defined to maximize the number
of counts and minimize the hardness ratio variation in each
interval. We fit the state--resolved spectra with a self--consistent
model for transmission, scattering and reflection of photons in
Compton--thick gas (\textsc{mytorus}). We assumed a clumpy
distribution for the obscuring and reflecting material with an
arbitrary effective global covering factor (i.e. the ``decoupled''
\textsc{mytorus} model). The model choice is driven by the observed
hard X--ray variability, strongly suggesting a low covering factor of
the obscuring matter in \textsc{NGC4945} (see e.g. Madejski et
  al. 2000, Done et al. 2003, Yaqoob et al. 2012). The
  state--resolved spectral analysis indicates that in all the states
  $A_{S90} << A_{Z90}$ and $A_{S0} \sim 20 \% \times A_{S90}$, which
  can be interpreted as a patchy absorber with a small covering factor
  and a large filling factor. This interpretation would be also
  consistent with the lack of strong column density variations, due to
  moving clouds.  A uniform distribution of obscuring gas (either a
  standard torus or a edge--on ring) are ruled out by the present
  observations.  The hard X--ray ($>$ 8~keV) components for the
  best--fit model are summarized in Fig. \ref{mod}. It is clear that
  most of the variability is due to the primary continuum, varying by
  a factor $\sim 3.3$.  The forward scattered component is variable,
  though with a lower amplitude (less than a factor 2), whereas the
  back--reflected component remain approximately constant within 10\%.
  Qualitatively the lower fractional variability of the
  scattered/reflected components is due to the larger
  path length of the reprocessed photons
  in the  obscuring medium. To roughly estimate the
  distance of this medium from the nucleus, we followed the method
  described in Marinucci et al. (2012). The circumnuclear
  gas is modelled as a cylinder with the axis on the plane of the sky,  radius
  R and height H. We assumed $H/R=0.13$ (i.e. the covering factor) and we
  considered a single {\it NuSTAR} $10-40$ keV light curve
  to determine the maximum observable flux variation of the reflected
  component as a function of the distance R. The fractional variability in the
  iron line flux is within  $\sim 5\%$.  The minimum distance needed
  to smear out the observed hard X--ray variability in the 10--40 keV
  energy range is  R$>10$~light days.  The weakly constrained lower
  limit on the distance of the reprocessor is due to the short
  timescales probed by continuous NuSTAR observations.  
  As a comparison, the  R$>35$~pc limit calculated by Marinucci et al. 2012 is obtained
  from the monitoring of the hard X--ray variability on a much longer time scale (i.e. the 65--month {\it Swift} BAT monitoring).

From a visual inspection of the broad--band best--fit spectra and
residuals in the four states (see Fig.~\ref{speratio}), there is no
evidence of a high energy curvature in the spectrum due to a cut--off
in the primary continuum.  The quality of the spectra is not
sufficient to break the degeneracy between the spectral
parameters. Assuming the power--law spectrum within the 90\% range of
the best--fit for each state, the addition of a high--energy
exponential roll-over does not improve the fit and the lower limits
(at the 90\% confidence level) on the e--folding energy are of the
order of $200-300$ keV in the L, M and H states. An exponential
cut--off at $E_C = 190^{+200}_{-40}$ keV improves the fit
($\Delta\chi^2 \sim 11$ for 1 degree of freedom) in the SH
state. Additional observations would be needed to confirm whether the
exponential cut--off is correlated with the source flux. The present
values agree fairly well with those previously reported by {\it
  BeppoSAX} ($\sim 80-500$ keV at the 90\% confidence level, Guainazzi
et al. 2000), {\it Suzaku} ($>80$ keV, Itoh et al. 2008) and {\it
    Swift} BAT ($>$ 100 keV, Yaqoob 2012).

We compute the $2-10$ keV intrinsic luminosity assuming the best--fit
parameters and correcting for the intrinsic absorption. The bolometric
luminosities in the L, M, H and SH states are calculated assuming the
bolometric corrections for type 2 AGN of Lusso et al. (2011). In the
range of the observed X--ray luminosities ($\sim 1.2- 5.4 \times
10^{42}$ erg s$^{-1}$) the bolometric corrections are of the order of
10 with a small dispersion. The inferred accretion rate $\lambda_{\rm
  Edd} = L_{\rm bol}/L_{\rm Edd}$, assuming a black hole mass of 1.4
$\times 10^{6} M_{\odot}$, is $\sim 0.07$, 0.14, 0.24 and 0.32 for the
L, M, S, and SH states, respectively. These values are significantly
smaller than those reported by Yaqoob (2012) from the analysis of a
large set of archival observations.  We note that Yaqoob (2012)
assumed a distance of about 8 Mpc (see his Tables~4 and 5), which is
$\sim$ twice that adopted in this paper; this explains the discrepancy
in the accretion rate estimates.

\begin{figure}
\centering
\includegraphics[width=8cm]{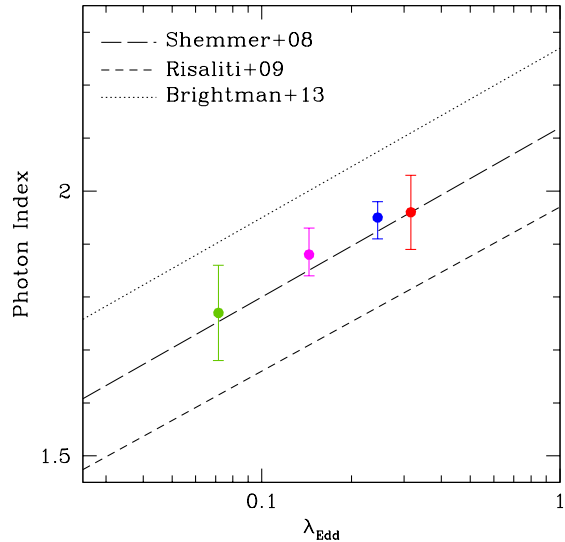}
\caption{Photon index versus the Eddington ratio for the L (green), M
  (magenta), H (blue) and SH (red) states. The long dashed, short
  dashed and dotted lines indicates the analytical fit found by
  Shemmer et al. (2008), Risaliti et al. (2009) and Brightman et
  al. (2013), respectively. }
\label{phedd}
\end{figure}

At face value, the accretion rate in the four different states
correlates with the intrinsic continuum spectral slope
(Fig. \ref{phedd}). However this relation is obtained wit the
assumption that the X--ray luminosity variability traces the
bolometric one in lockstep and with the same fractional degree of
variability. This assumption may not necessarily be true and it is
known that variability timescales are usually longer in the UV than in
the X--rays (i.e. Maoz et al. 2002, Nandra et al. 2000, Clavel et
al. 1992).

Even if the slope of the relation and its normalization are close
to that of a few AGN type 1 samples for which the accretion rate could be
measured (Shemmer et al. 2008, Risaliti et al. 2009, Brightman et
al. 2013), we caution about the use of single--epoch X--ray data to
derive $\lambda_{\rm Edd}$ values for obscured AGN in the distant
Universe.

\begin{figure*}
\begin{tabular}{cccc}
\centering
\includegraphics[width=3.7cm,height=4cm]{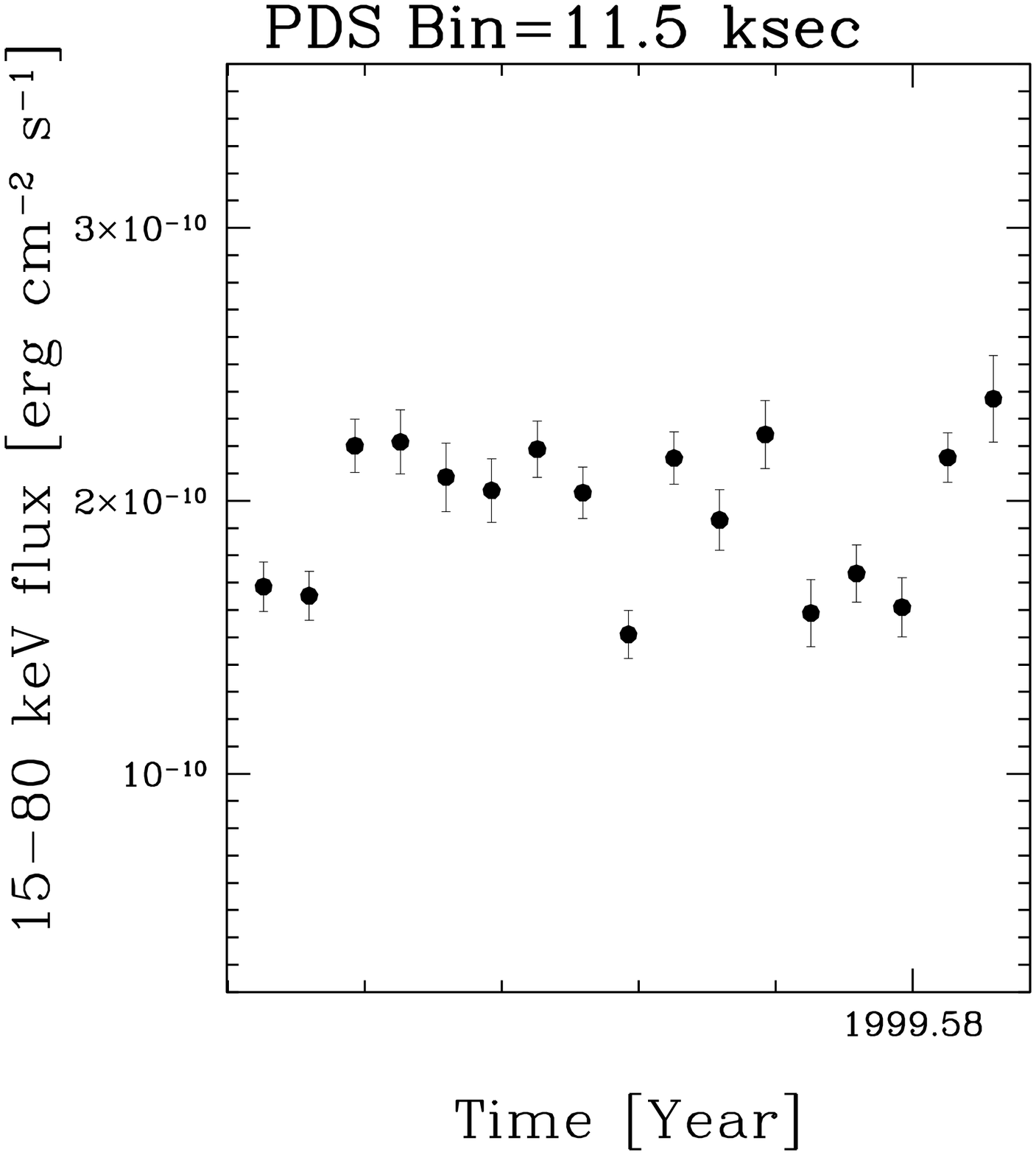}
\includegraphics[width=4.4cm,height=4cm]{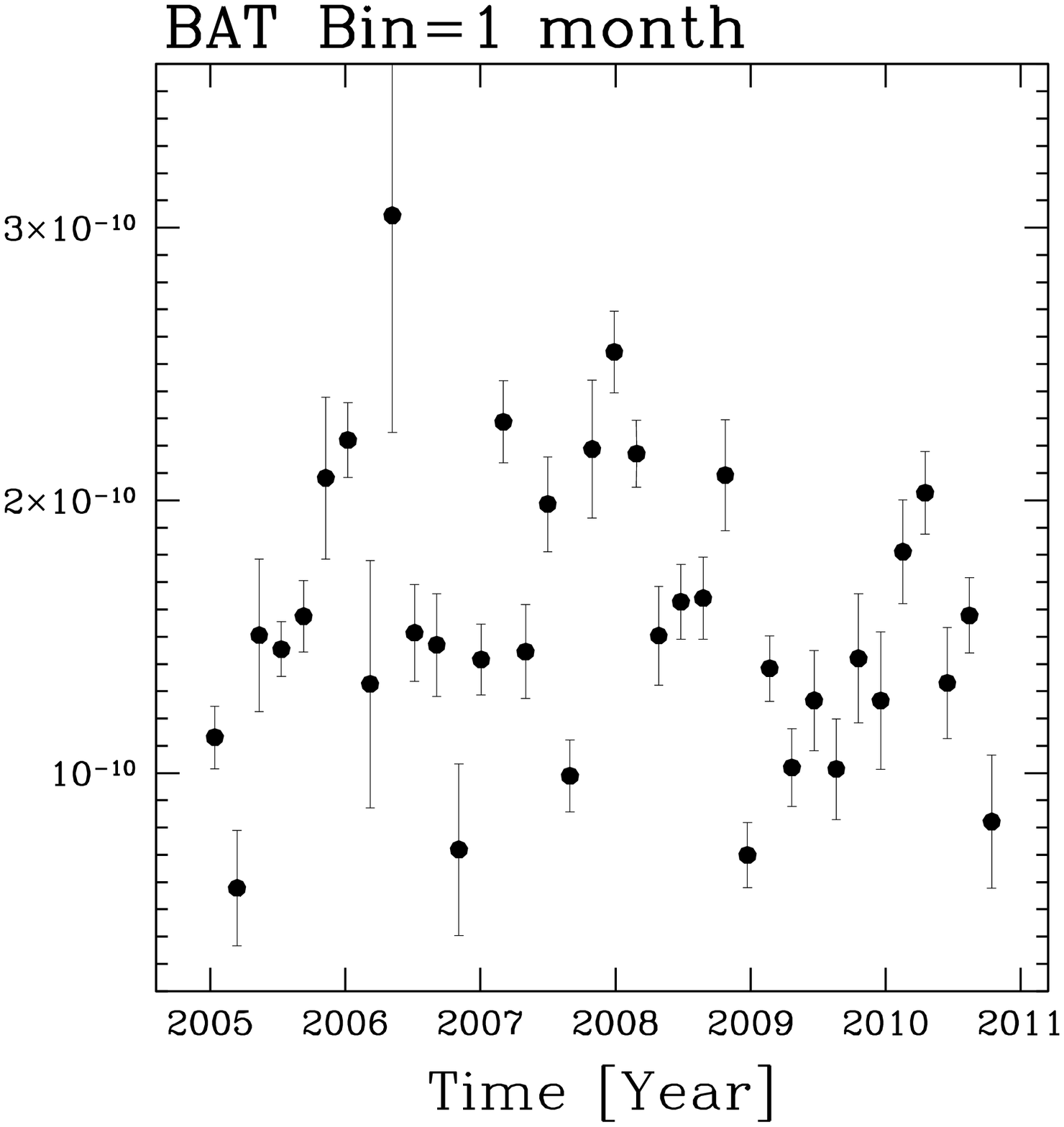}
\includegraphics[width=6cm,height=4cm]{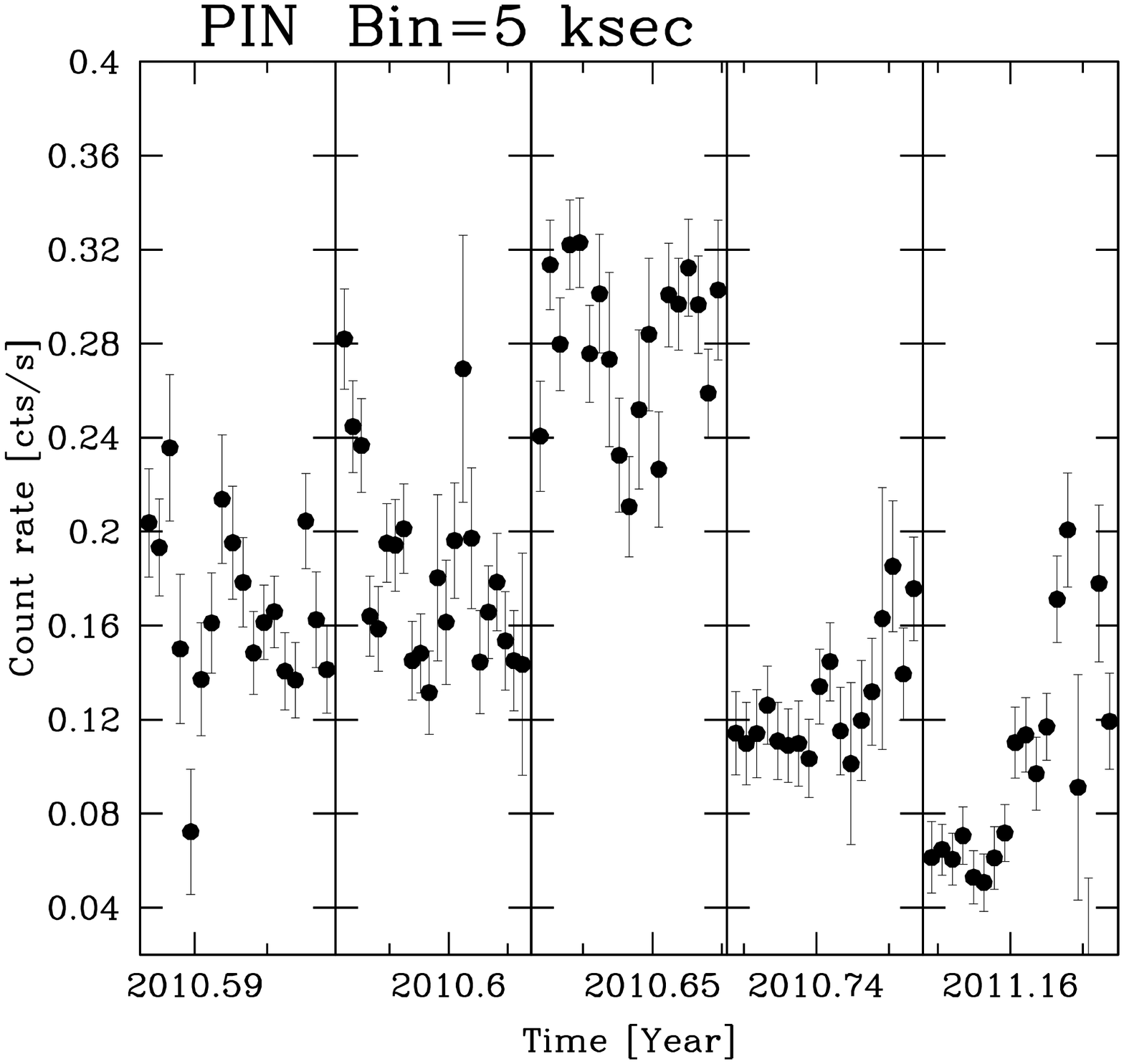}
\includegraphics[width=4.6cm,,height=4cm]{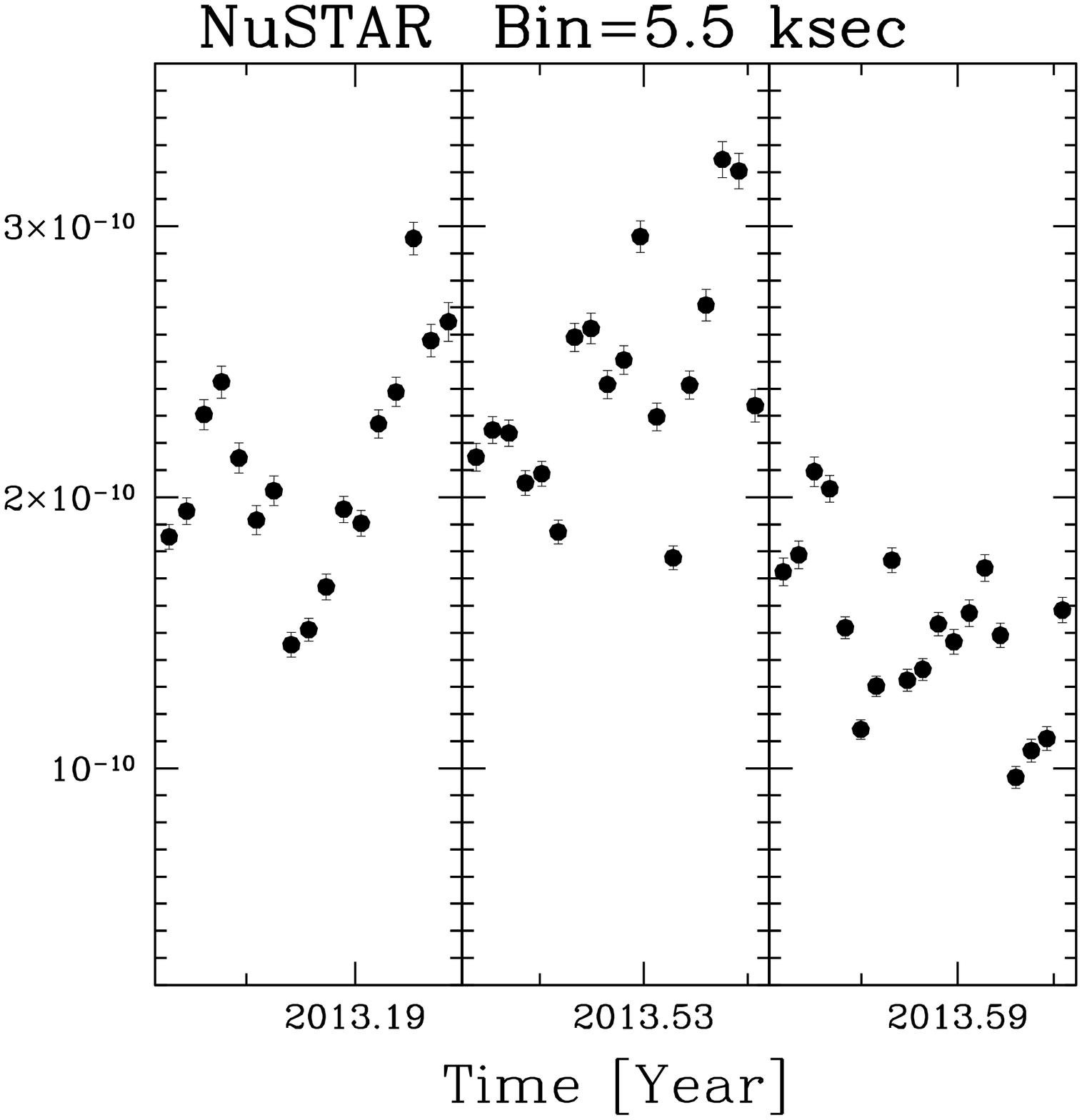}
\end{tabular}
\caption{From right to left panel the {\it BeppoSAX} PDS, {\it Swift}
  BAT, {\it Suzaku} HXD PIN and $15-79$ keV {\it NuSTAR}
  background--subtracted light curves in bins of 11.5~ksec, 1~month,
  5~ksec and 5.5 ksec, respectively. The observed count rates are
  converted to the $15-80$ keV X--ray flux assuming the average
  best--fit values from the {\it NuSTAR} spectra (see \S~6).}
\label{lcurveshist}
\end{figure*}

The most remarkable and unique characteristic of the \textsc{NGC4945}
hard ($>$ 10 keV) X--ray emission is the large amplitude and fast
variability of the primary continuum piercing through a constant
column density Compton thick absorber. In Fig.~\ref{lcurveshist} we
reported the history of hard X--ray variability from {\it BeppoSAX}
PDS, {\it Swift} BAT, {\it Suzaku} PIN and {\it NuSTAR} data. The
observed count rates are converted to the $15-80$ keV X--ray flux
assuming the average best--fit values from the {\it NuSTAR} spectra
($\Gamma=1.9$ and $N_H = 3.55 \times 10^{24}$ cm$^{-2}$). The
unprecedented statistical quality of the {\it NuSTAR} observations is
evident from a visual inspection of the panel. With the possible
exception of {\it BeppoSAX} PDS, the amplitude variability is similar
among the various observations and fully sampled by {\it NuSTAR} data.
The observed range of accretion rates is thus likely to be
representative of the nuclear accretion history over the last $\sim
15$ years.  \par The nucleus of \textsc{NGC4945} is accreting at a
rate which is faster ($\sim 0.1-0.3$) than the typical values
$\lambda_{\rm Edd} < 0.1$ of obscured Seyfert 2 galaxies in the local
Universe (Vasudevan et al. 2010). The possibility of episodes of super
Eddington accretion over the past $\sim15$ years, are ruled out by the
present analysis.  Even though not as extreme as previously claimed,
the typical accretion rate of \textsc{NGC4945} lies on the extreme of
the distribution observed for Seyfert 2 galaxies and is more typical
of a luminous QSO.

The low covering factor of
the Compton thick obscuring gas, as inferred for the first time on a
sound statistical footing via state--resolved spectral analysis, and the
high accretion rate make \textsc{NGC4945} an extremely interesting
source. The observed properties are typical of the AGN population
postulated in the synthesis models for the X--ray background to fit
the $20-30$ keV peak and expected to be abundant at $z\simeq 1-2$.
Further X--ray monitoring of this ``Rosetta Stone'' would further constrain
the geometry and the physics of the most
obscured, rapidly accreting black holes. More specifically, a dense
monitoring would allow a finer analysis  of the state--resolved
spectral analysis. Moreover, long term variability would place stronger
constraints on the distribution of the absorbing material.

\section{Conclusion}

We present the {\it NuSTAR} spectra of \textsc{NGC4945}, one of the
brightest Seyfert 2 galaxies in the local universe.  Three {\it
  NuSTAR} observations obtained over a time interval of about 5 months
provide spectra and variability of unprecedented quality, allowing detailed modelling of
the source geometry.  The primary findings can be summarized as follows:

\begin{itemize}

\item in the {\it NuSTAR} band the source is highly variable with a
  doubling/halving time as short as 16 ksec. The maximum variability
  is in the $10-40$ keV band, while below 10 keV the source is
  constant.

\item  The analysis of the hardness ratio light curves implies strong
  spectral variability above 8--10 keV, which has been addressed for
  the first time in this source, via state--resolved spectral analysis
  of the {\it NuSTAR} high energy spectra.

\item The state--resolved spectral analysis, confirmed that the
  constant X--ray flux below 10 keV is consistent with being due to
  reflection from Compton--thick material. The shape of the reflected
  continuum below 10 keV and the iron line complex has been determined
  by combining {\it NuSTAR} with {\it Suzaku} and {\it Chandra}
  observations. 

\item The variability in the hard X--ray ($>$ 10 keV) band is mainly
  due to variation in the intensity of the primary continuum piercing
  through a high column density $N_{\rm H}$ $\sim 3.6 \times$
  10$^{24}$ cm$^{-2}$ absorber. There is evidence of a steepening of
  the primary continuum spectral slope with increasing flux. The
  spectral and variability properties indicate, in agreement with
  previous results, a small covering factor ($\sim 0.1-0.15$) for the
  obscuring matter which is best parameterized in terms of a clumpy
  toroidal distribution.

\item A power spectral density analysis suggests that the variability
  is almost identical to that observed in Seyfert 1 galaxies,
  reinforcing the interpretation of spectral variability as due to
  variations in the primary emission.

\item There is no clear evidence for a high--energy cut--off with
  lower limits of the order of $\sim 200-300$ keV.

\item The intrinsic continuum variability is associated with a
  variation of the Eddington ratio in the range $0.1-0.3$. Assuming a
  distance of $\sim$ 3.8 Mpc there is no evidence of super Eddington
  accretion.

\item With the assumption that the X--ray luminosity variations
   trace the bolometric luminosity, the accretion rate  correlates with the intrinsic spectral
  index, in agreement with the trend observed for relatively
  high--redshift AGN. The fact that the same trend is observed for
  different states of the same source lend further support to a close
  link between the hard X--ray slope and the physics of accretion processes.
  
\end{itemize}

\acknowledgements 

This work was supported under NASA Contract NNG08FD60C, and made use
of data from the {\it NuSTAR} mission, a project led by the California
Institute of Technology, managed by the Jet Propulsion Laboratory, and
funded by the National Aeronautics and Space Administration. We thank
the {\it NuSTAR} Operations, Software and Calibration teams for
support with the execution and analysis of these observations. This
research has made use of the {\it NuSTAR} Data Analysis Software
(NuSTARDAS) jointly developed by the ASI Science Data Center (ASDC,
Italy) and the California Institute of Technology (USA). SP, AC, FF
and GM acknowledge support from the ASI/INAF grant I/037/12/0 --
011/13. AC acknowledges the Caltech Kingsley visitor program. PG
acknowledges support from STFC (grant reference ST/J003697/1). GBL
acknowledges support from STFC (grant reference ST/K501979/1). DMA
acknowledges support from STFC (grant reference ST/I001573/1) and form
the Leverhulme Trust. PA acknowledges financial support from Fondecyt
grant 11100449 and Anillo ACT1101. GR acknowledges financial support
from grant NASA GO3--14109X. MK gratefully acknowledges support from
Swiss National Science Foundation Grant PP00P2\_138979/1. WNB and BL
acknowledge support from California Institute of Technology NuSTAR
subcontract 44A--1092750 and NASA ADP Grant NNX10AC99G. FEB
acknowledges support from Basal--CATA PFB--06/2007, CONICYT--Chile
(grants FONDECYT 1101024 and ``EMBIGGEN'' Anillo ACT1101), and Project
IC120009 ``Millennium Institute of Astrophysics (MAS)'' of Iniciativa
Cient\'{\i}fica Milenio del Ministerio de Econom\'{\i}a, Fomento y
Turismo. SP is grateful to Tahir Yaqoob for useful discussions
on the \textsc{mytorus} model.


\end{document}